\documentclass[draftclsnofoot,12pt,onecolumn]{IEEEtran}
\usepackage{tikz}
\usetikzlibrary{positioning,shapes,shapes.geometric,arrows}
\usepackage{pgfplots}
\pgfplotsset{compat=newest}

\ifCLASSINFOpdf

   \usepackage{subfigure}

\else
\fi
\usepackage[cmex10]{amsmath}
\usepackage{amssymb}
\usepackage{bbold}
\usepackage{dsfont}
\usepackage{bm}
\usepackage{cite}

\usepackage{algorithmic}
\usepackage{algorithm}
\begin{document}
\title{Licensed and Unlicensed Spectrum Allocation in Heterogeneous Networks}

\author{\IEEEauthorblockN{Zhiyi Zhou, Dongning Guo, and Michael L. Honig\\}
\IEEEauthorblockA{Department of Electrical Engineering and Computer Science\\ Northwestern University, Evanston, IL 60208, USA}

\thanks{This work was supported in part by a gift from Futurewei Technologies
and by the National Science Foundation under Grant No. CCF-1423040.}
}


%Telephone: (800) 555--1212, Fax: (888) 555--1212}


\maketitle

\begin{abstract}
%\boldmath
In future networks, an operator may employ a wide range of access points using diverse radio access technologies (RATs) over multiple licensed and unlicensed frequency bands. This paper studies centralized user association and spectrum allocation across many access points in such a heterogeneous network (HetNet). Such centralized control is on a relatively slow timescale to allow information exchange and joint optimization over multiple cells. This is in contrast and complementary to fast timescale distributed scheduling. A queueing model is introduced to capture the lower spectral efficiency, reliability, and additional delays of data transmission over the unlicensed bands due to contention and/or listen-before-talk requirements. Two optimization-based spectrum allocation schemes are proposed along with efficient algorithms for computing the allocations. The proposed solutions are fully aware of traffic loads, network topology, as well as external interference levels in the unlicensed bands. Packet-level simulation results show that the proposed schemes significantly outperform orthogonal and full-frequency-reuse allocations under all traffic conditions.
\end{abstract}


\begin{IEEEkeywords}
Long term evolution (LTE), LTE in unlicensed spectrum (LTE-U), multiple radio access technologies, spectrum allocation, user association.
\end{IEEEkeywords}

% \fi
\IEEEpeerreviewmaketitle

\section{Introduction}
The wireless industry has set an ambitious goal to increase the area capacity (in bits per second per square meter) by three orders of magnitude in the next five to ten years. In addition to densely deploying small cells and improving the spectral efficiency \cite{LTE:Khan}, another avenue is to exploit all available spectrum, which is a relatively scarce resource \cite{What:Andrews}. Future generation cellular networks are likely to involve multiple radio
access technologies (RATs) over multiple frequency bands (including millimeter wave). Such RATs include Long Term Evolution (LTE), WiFi, and LTE in unlicensed spectrum (LTE-U). The prime bands today are licensed frequency bands under 3 GHz and over 500 MHz of unlicensed spectrum in the 2.4 GHz and 5 GHz frequency bands.

Current 4G cellular networks generally use regular frequency reuse patterns, including full frequency reuse and fractional frequency reuse.  In the former scheme, every cell uses all available frequency bands, whereas the latter scheme is similar, except that cells use orthogonal frequency bands at the cell edge to reduce mutual interference. However, such simple methods are not as effective in emerging heterogeneous networks (HetNets), whose topologies are often highly irregular, and whose small cells have widely varying traffic conditions.

There have been many studies on spectrum allocation in cellular networks (see, e.g., \cite{Etkin:Spectrum,Joint:Huang,Stolyar:Self,Chang:Multicell,Ali:Dynamic,Madan:Cell,Liao:Base,Joint:Fool,Energy:Lim,Distributed:Shen,Algorithms:Deb,Energy:Kuang,Kuang:Optimal}). Most authors formulate the allocation problem as that of deciding, for each slice of the spectrum, which access points (APs) and/or links should use it. In addition, user association was considered in \cite{Joint:Huang,Joint:Fool}. In general, NP-hard nonconvex integer programming problems are formulated, which typically have many local optima. Furthermore, the figures of merit used in most work are physical layer performance measures such as sum rate and outage probability. These traffic-independent metrics may not reflect a user's relevant quality of service (QoS) in HetNets with large traffic variations in overlapping cells with complicated interference conditions.

To address the preceding issues, an alternative optimization-based framework was developed in \cite{Traffic:Zhuang} and \cite{Energy:Zhuang} for allocating downlink spectrum resources in a HetNet. This framework allows arbitrary user association and flexible spectrum allocation, driven entirely by users' traffic loads. This framework is well matched to emerging centralized remote processing architectures such as cloud radio access network (C-RAN) and cell-free massive multiple-input multiple-output systems \cite{Cell:Ngo}. In contrast to most existing work \cite{Stolyar:Self,Chang:Multicell,Ali:Dynamic,Madan:Cell,Liao:Base}, which considers resource allocation on the timescale of a frame, here the timescale of resource adaptation is conceived to be once every few seconds or minutes. This timescale is, on the one hand, fast enough for tracking aggregate traffic variations and large-scale fading, and, on the other hand, slow enough to allow information exchange and joint optimization of many APs with a large number of user equipments (UEs). Another advantage is that the spectrum resources can be assumed to be homogeneous on this timescale, namely, every segment in the same band has about the same spectral efficiency when averaged over many frames. The approach in \cite{Traffic:Zhuang} has been generalized to incorporate energy-efficient allocation via cell activation \cite{Energy:Zhuang} as well as the effect of opportunistic scheduling on a fast timescale \cite{teng2015resource,teng2015Dual}.

In this work we generalize the framework of \cite{Traffic:Zhuang} and \cite{Energy:Zhuang} to the scenario in which there are multiple RATs over multiple frequency bands. Resource allocation over multiple RATs has been studied in \cite{Small:Liu,Resource:Elsherif,Energy:Chen}. In \cite{Small:Liu}, the authors presented a scheme for balancing licensed and unlicensed traffic in the case of a single femto user and single WiFi user. In \cite{Resource:Elsherif,Energy:Chen}, joint licensed and unlicensed resource allocation algorithms were proposed for licensed-assisted access systems for throughput and energy efficiency maximization, respectively. However, the distinct characteristics of the unlicensed bands were not modeled in those papers. In our work, different queueing models are proposed to distinguish the characteristics of licensed and unlicensed bands. The multiple-band allocation problem is formulated as a bi-convex optimization problem. A conservative allocation scheme is first proposed following the approach of \cite{Traffic:Zhuang} and \cite{Energy:Zhuang}. This is followed by a utilization-dependent allocation scheme which incorporates the utilizations of the APs into the formulation to more accurately account for dynamic inter-cell interference. An iterative algorithm is designed to solve the allocation problem with manageable computational complexity for small systems. In each step, we solve a convex problem with a unique optimal solution.

Another feature of this work is that the packet length is allowed to have a general probability distribution. In prior work \cite{Energy:Zhuang,Traffic:Zhuang,teng2015resource} and \cite{teng2015Dual}, the packet length is assumed to be exponentially distributed to yield a simple analytic form in the objective representing the QoS. We show that the proposed formulation and algorithm apply to general packet length distributions and hence a broader class of traffic conditions.

To evaluate the performance, packet-level simulations are carried out. It is shown that both the conservative allocation scheme and the utilization-based allocation scheme significantly reduce the average packet delay in the heavy traffic regime compared to orthogonal allocation and full-frequency-reuse allocation. The large performance gain is observed mainly because the proposed schemes are traffic-aware and also exploit the particular characteristics of each RAT. In addition, the utilization-dependent allocation scheme attains the best performance over all traffic conditions due to its accurate modeling of the dynamic interference among APs.

One practical application of this work is the emerging C-RAN which allows many remote radio function units connect to a centralized network controller. The total overhead for the network controller to perform the proposed resource allocation scheme includes collecting the spectral efficiencies of all links and users' traffic information. Since the timescale of resource adaptation is considered to be once every a few seconds or minutes, the overhead is quite small. For example, the rate for sending 30,000 parameters (16 bits each) every minute is only 8 kilobits per second (kbps).

The remainder of this paper is organized as follows. The system model is introduced in Section II. An optimization problem using the conservative allocation is formulated in Section III. A utilization-dependent allocation scheme is presented in Section IV. The extension to general packet length distributions is given in Section V. Simulation results are presented in Section VI and concluding remarks are given in Section VII. All technical proofs are relegated to the appendices.

\section{System Model}
In this section, we introduce models for user traffic, spectrum allocation, and link throughput. The models extend those in \cite{Energy:Zhuang} to accommodate multiple RATs. A new queueing model is then introduced for traffic over unlicensed bands.
\subsection{Spectrum Allocation}
We consider the downlink of a heterogeneous network consisting of $n$ APs and many UEs. Without loss of generality, suppose each AP can use all $m$ different RATs, with each RAT on a separate frequency band. Since UEs located near each other often have similar channel conditions, these UEs can be treated as a group on the slow timescale. This is a generalization of the extreme case where each UE group contains only one single UE. Moreover, it suffices to carry out the slow timescale allocation to the UE groups rather than individual UEs.

Denote the set of all APs by $\mathcal{N}=\{1,...,n\}$, the set of all user groups by $\mathcal{\mathcal{K}}=\{1,...,k\}$, and the set of RATs by $\mathcal{M}=\{1,...,m\}$. RAT $l$ employs its separate homogeneous spectrum of bandwidth $w^{(l)}$. We allow arbitrary association so that each AP can simultaneously serve any subset of UE groups and each UE can be simultaneously served by any subset of APs. Furthermore, we allow flexible resource allocation in that each AP-UE link can use an arbitrary number of RATs, where each RAT uses an arbitrary (possibly discontinuous) subset of the available spectrum. Despite the enormous number of possibilities, we will show that the actual AP-UE association and spectrum allocation is extremely sparse.

\begin{figure}
	\centering
\begin{tikzpicture}
\begin{axis}[
xbar stacked,
ytick=data,
y axis line style= { draw opacity=0 },
axis y line*=none,
axis x line*=bottom,
xtick={0,1},
width=.897\textwidth,
bar width=6mm,
yticklabels={AP3,AP2,AP1},
xmin=0,
xmax=1,
% area legend,
y=8mm,
enlarge y limits={abs=0.625},
]
\addplot[white,fill=white] coordinates
{(0.1,0) (0.1,1) (0,2)};
\addplot[white,fill=white] coordinates
{(0.15,0) (0,1) (0,2)};
\addplot[orange,fill=orange] coordinates
{(0,0) (0,1) (0.1,2)};
\addplot[red,fill=red] coordinates
{(0,0) (.15,1) (.15,2)};
\addplot[green,fill=green] coordinates
% Transfer
{(0.2,0) (0.2,1) (.2,2)};
\addplot[yellow,fill=yellow] coordinates
{(.25,0) (0,1) (.25,2)};
\addplot[white,fill=white] coordinates
{(0,0) (0.25,1) (0,2)};
\addplot[cyan,fill=cyan] coordinates
{(0,0) (0.075,1) (0,2)};
\addplot[white,fill=white] coordinates
{(0.075,0) (0,1) (0.075,2)};
\addplot[magenta,fill=magenta] coordinates
{(0.125,0) (0.125,1) (0,2)};
\addplot[blue,fill=blue] coordinates
{(0.1,0) (0,1) (0,2)};
\end{axis}%
\node at (2.33,-.3) {$y^{\{1,2\}}$};
\draw (1.33,-.2) -- (1.33,2.5);
\node at (0.67,-.3) {$y^{\{1\}}$};
\draw (3.31,-.2) -- (3.31,2.5);
\node at (4.63,-.3) {$y^{\{1,2,3\}}$};
\draw (5.95,-.2) -- (5.95,2.5);
\node at (7.61,-.3) {$y^{\{1,3\}}$};
\draw (9.27,-.2) -- (9.27,2.5);
\node at (9.77,-.3) {$y^{\{2\}}$};
\draw (10.27,-.2) -- (10.27,2.5);
\node at (11.09,-.3) {$y^{\{2,3\}}$};
\draw (11.92,-.2) -- (11.92,2.5);
\node at (12.62,-.3) {$y^{\{3\}}$};
<5->{
	\draw[dashed] (7.71,0.12) -- (7.71,.96);
	\draw[dashed] (7.35,1.75) -- (7.35,2.5);
	\node at (6.75,2.1) {\small $x_{1\rightarrow 1}^{\{1,3\}}$};
	\node at (8.35,2.1) {\small $x_{1\rightarrow 2}^{\{1,3\}}$};
	\node at (6.95,.5) {\small $x_{3\rightarrow 1}^{\{1,3\}}$};
	\node at (8.5,.5) {\small $x_{3\rightarrow 2}^{\{1,3\}}$};
}
\end{tikzpicture}
	\caption{Illustration of all patterns of a 3-AP network. The allocations to 2 users are revealed under pattern $\{1,3\}$.}
	\label{fig1}
\end{figure}
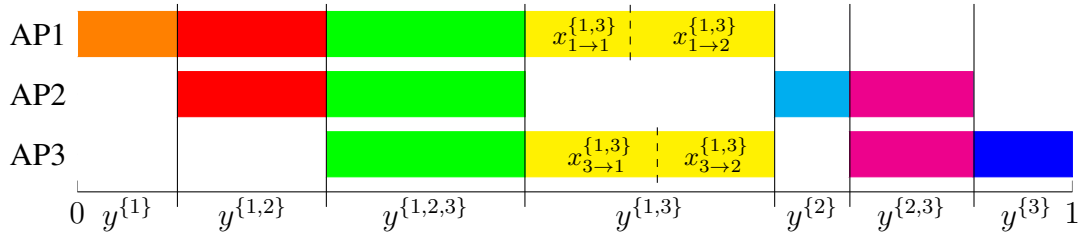

The key to total spectrum agility is the notion of {\em pattern} \cite{Traffic:Zhuang,Energy:Zhuang}. In general, a pattern simply refers to a subset of transmitters. A time-frequency resource is said to be reserved for pattern $\mathcal{A}$ if the resource is to be shared by transmitters in $\mathcal{A}$.\footnote{The notion of pattern finds its root in the concept of independent set defined in the special case where the network is described by a conflict graph, and where nodes/links in the independent set share the same resources \cite{R:Com}.} In the downlink, a pattern $\mathcal{A}$ is a subset of $\mathcal{N}$, and all APs in $\mathcal{A}$ have access to the time-frequency resources associated with pattern $\mathcal{A}$. Assuming known transmit power spectral densities (PSDs), the pattern of a resource determines the signal-to-interference-plus-noise ratio (SINR) and hence the spectral efficiencies of all links over the resource. The allocation problem can then be formulated as how to divide the spectrum of each RAT among all $2^n-1$ nonempty patterns. To illustrate the concept of pattern with an example as shown in Fig. \ref{fig1}, suppose there are three APs operating over one frequency band. The spectrum can be divided into $2^3-1=7$ segments, where one segment is used by AP 1 exclusively (the pattern is $\{1\}$), a second is used by AP 2 exclusively (the pattern is $\{2\}$), a third is used by AP 3 exclusively (the pattern is $\{3\}$) and the remaining four segments include three shared by the two APs (the patterns are $\{1,2\}$, $\{2,3\}$, and $\{1,3\}$, respectively) as well as one segment shared by all three APs (the pattern is $\{1,2,3\}$). If the transmit pattern of a certain spectrum resource is $\{1,2\}$, i.e., both APs transmit, then the spectral efficiencies of all links are determined as discussed below.

A network controller collects traffic load and channel/interference information from all the APs. Because the time period of slow timescale allocation is much longer than the channel coherence time, the channel conditions are modeled using path loss and the statistics of fading. Given the average channel and traffic conditions, the task of the central controller is to determine which spectrum segments to allocate to each AP, and subsequently, which sub-segments to allocate to specific UEs associated with that AP. To be precise, we need to solve the following three subproblems:
\begin{enumerate}
\item Decide which RAT or RATs should be used to serve each UE group.
\item Allocate the bandwidth across all $2^n$ patterns for each RAT. This is denoted by a $2^n \times m$-tuple: $y = (y^{\mathcal{A},l})_{\mathcal{A} \subset \mathcal{N},l \in \mathcal{M}}$, where $y^{\mathcal{A},l} \in [0, 1]$ is the fraction of spectrum assigned to RAT $l$ shared by APs in set $\mathcal{A}$. Clearly,
\begin{equation}
\sum_{\mathcal{A}\subset \mathcal{N}}y^{\mathcal{A},l}=1,\ \ \ \ \forall l\in \mathcal{M}
\end{equation}
and any efficient allocation would not use the empty pattern, so that $y^{\varnothing,l} = 0$, for every $l\in \mathcal{M}$.
\item For every pattern $\mathcal{A}\subset\mathcal{N}$, for every AP $i\in\mathcal{A}$, divide the bandwidth $y^{\mathcal{A},l}$ to the UE groups. Denote the bandwidth allocated to the link $i \rightarrow j$ over pattern $\mathcal{A}$ under RAT $l$ as $x_{i \rightarrow j}^{\mathcal{A},l}$. As shown in Fig. 1, $y^{\{1,3\}}$ (colored yellow) is further divided by AP 1 into two parts to serve UE group 1 and UE group 2, respectively, while AP 3 divides the same shared spectrum differently. Then we have:
\begin{equation}
\sum_{j \in \mathcal{K}}x_{i \rightarrow j}^{\mathcal{A},l} \leq y^{\mathcal{A},l}, \ \ \ \ \forall i \in \mathcal{A}, \mathcal{A}\subset \mathcal{N}, l\in \mathcal{M}.
\end{equation}
\end{enumerate}
The user association is indirectly determined by the amount of spectrum resources assigned by each AP to each group. Specifically, UE $j$ is assigned to AP $i$ over RAT $l$ if and only if $x_{i \rightarrow j}^{\mathcal{A},l}>0$ for some pattern $\mathcal{A}$.

\subsection{From Spectrum to Transmission Rates}
An important measure of user QoS is the average packet delay in the system. The delay is determined by the packet arrival statistics and the service rates. The instantaneous service rate of a UE group's queue in turn depends on the spectral efficiencies along with the bandwidths allocated to that specific UE group. For simplicity, it is assumed that when AP $i$ transmits over RAT $l$, it employs all patterns available to it and applies a flat PSD $p^{(l)}_i$ over the allocated spectrum. At any frequency designated for RAT $l$, the instantaneous spectral efficiency achievable by the link from AP $i$ to UE $j$ depends on the set of active APs $\mathcal{A}\subset \mathcal{N}$ using that frequency. Let this (spectral) efficiency be denoted by $s_{i \rightarrow j}^{\mathcal{A},l}$. Clearly, $s_{i \rightarrow j}^{\mathcal{A},l}=0$ if $i\notin \mathcal{A}$. Moreover, the spectral efficiency decreases as more APs become active, i.e., $s_{i \rightarrow j}^{\mathcal{A},l} \geq x_{i \rightarrow j}^{\mathcal{B},l}$ if $i\in \mathcal{A}\subset \mathcal{B}$. On the slow timescale, the spectral efficiencies are either known \textit{a priori} or can be measured and sent to the central controller. For later convenience, we normalize the spectral efficiency over RAT $l$ using factor $L/w^{(l)}$ (bits/second/Hz) so that the unit of $s_{i \rightarrow j}^{\mathcal{A},l}$ is packets/second.

For concreteness in obtaining numerical results, we use Shannon's formula to arrive at the expression for link efficiencies:
\begin{equation}
s_{i \rightarrow j}^{\mathcal{A},l}=\frac{\alpha^{(l)} w^{(l)}\mathbb{1}(i\in \mathcal{A})}{L}\log_{2}\left(1+\frac{p^{(l)}_i}{I_{i\rightarrow j}^{\mathcal{A},l}}\right)\ \ \textrm{packets/s},
\label{shannon}
\end{equation}
where $L$ is the average packet length in bits,\footnote{Packet lengths are assumed to be i.i.d. with exponential distribution; later we will show that the proposed methods can be applied to general packet length distributions.} $\alpha^{(l)}\in (0,1]$ is the discount factor for RAT $l$, $\mathbb{1}(i\in \mathcal{A}) = 1$ if $i \in \mathcal{A}$ and $\mathbb{1}(i\in \mathcal{A}) = 0$ otherwise, and $I_{i\rightarrow j}^{\mathcal{A},l}$ is the total noise plus interference PSD from other APs in $\mathcal{A}$ to UE $j$, which depends on their transmit PSDs and path loss. The discount factor of a licensed band is typically closer to 1 than that of an unlicensed band due to external interference from other operators in the unlicensed band. The discount factor is not crucial and is merely included for flexibility of the model. The average effect of small-scale fading can be included by considering the ergodic capacity in lieu of (\ref{shannon}), which does not change the main developments of this paper.

Over a resource reserved for pattern $\mathcal{A}$, only the subset of APs in $\mathcal{A}$ with data to transmit will be using the resource at any given time. In general, the instantaneous service rate depends on the set of transmitting APs and the rate adaptation scheme. Two service rate models are adopted as described in
the following.

Under the conservative allocation introduced in \cite{Traffic:Zhuang}, if some of RAT $l$'s spectrum resources reserved for pattern $\mathcal{A}$ are allocated to link $i\rightarrow j$, then AP $i$ transmits UE $j$'s packets at spectral efficiency $s_{i\rightarrow j}^{\mathcal{A},l}$ using RAT $l$ over pattern $\mathcal{A}$. This rate is achievable even if all APs in $\mathcal{A}$ have data to transmit (hence the name conservative). This is equivalent to assuming that other APs' traffic is always backlogged. The rate contributed by spectrum reserved for pattern $\mathcal{A}$ under RAT $l$ is the product of the spectral efficiency and the bandwidth: $s_{i\rightarrow j}^{\mathcal{A},l}x_{i\rightarrow j}^{\mathcal{A},l}$. The total service rate for the queue of UE $j$ over RAT $l$ is the sum rate of all APs and patterns, expressed as:
\begin{equation}
r^{(l)}_j=\sum_{\mathcal{A}\subset \mathcal{N}}\sum_{i\in \mathcal{N}}s_{i\rightarrow j}^{\mathcal{A},l}x_{i\rightarrow j}^{\mathcal{A},l} \ \ \textrm{packets/s}.
\label{eq4}
\end{equation}
The advantage of the conservative allocation is that there is no need for the scheduler (or whichever unit that performs rate adaptation) to know the state of the other APs included in the pattern. However, the conservative rate (\ref{eq4}) is a lower bound on the actual rate because the interference is overestimated.

To adapt the allocation to the actual interference pattern, we introduce a new model in this work, referred to as the {\em utilization-dependent model}. Under this model, the service rate contributed by the spectrum reserved for pattern $\mathcal{A}$ depends on the subset of transmitting APs in $\mathcal{A}$. Specifically, the service rate for UE group $j$ over RAT $l$ when the set of active interfering APs is $\mathcal{I}$ is expressed as:
\begin{equation}
r^{\mathcal{I},l}_j=\sum_{\mathcal{A}\subset \mathcal{N}}\sum_{i\in \mathcal{N}}s_{i\rightarrow j}^{\mathcal{A}\cap \mathcal{I},l}x_{i\rightarrow j}^{\mathcal{A},l} \ \ \textrm{packets/s}.
\end{equation}

\subsection{Queueing Model}

Without loss of generality, we restrict the subsequent investigation to the scenario with two RATs, where one RAT is over the licensed band (conceived as LTE) and the other is over the unlicensed band (conceived as LTE-U). This can be easily generalized to more than two RATs.

The traffic to UE group $j$ is modeled as an independent Poisson point process with rate $\lambda_{j}$ packets per second. The packet lengths are independent random variables whose average is $L$ bits. Packets intended for each UE group are transmitted according to the first-in-first-out (FIFO) discipline. The buffer in each queue is assumed to be unlimited for simplicity. The traffic load of UE group $j$ is further divided into two streams served by the two RATs, respectively. This may be implemented by dividing the group of UEs for association with different RATs, so that each UE is only served by one RAT. The resulting packet arrival rates of two streams are denoted as $\lambda^{(1)}_j$ and $\lambda^{(2)}_j$, respectively, which are variables to be optimized subject to the constraint: 
\begin{equation}
\lambda^{(1)}_j+\lambda^{(2)}_j \geq \lambda_j.
\end{equation}
For each RAT, the corresponding $k$ UE groups form a system of $k$ interactive queues where the instantaneous service rate of each queue may depend on which other queues are empty.

The physical and/or medium access control layers of LTE-U are designed to facilitate coexistence with other RATs in unlicensed bands, such as WiFi.\footnote{Challenges of such coexistence are discussed in~\cite{jindal2015lte}.} In \cite{License:Rapeepat}, a listen-before-talk scheme was proposed in which carrier sensing is embedded in a deterministic portion of a LTE subframe. In \cite{LTE_UL:Fabiano}, the authors proposed the use of LTE uplink power control to improve LTE/WiFi coexistence. References \cite{Enabling:Almeida} and \cite{System:Timo} propose to enable LTE/WiFi coexistence by muting LTE transmission on certain subframes following a pre-determined pattern.

\begin{figure}[t]
	\centering
	\includegraphics[width=5in]{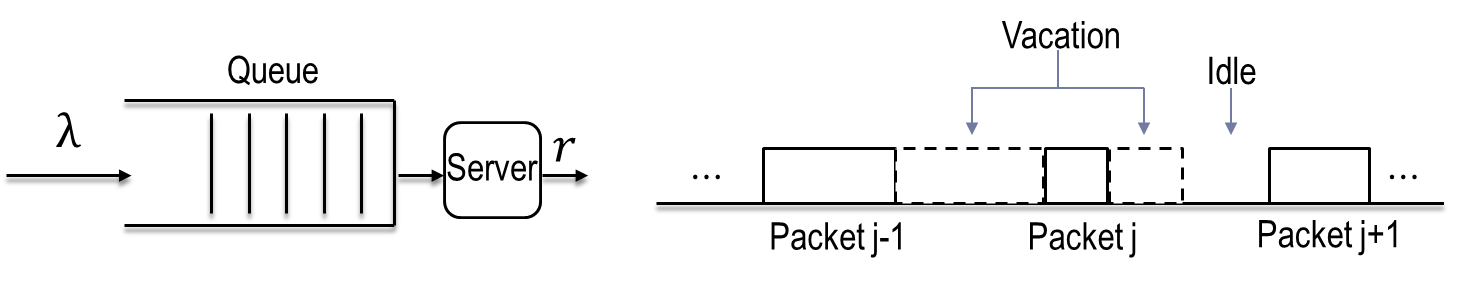}
	\caption{Illustration of queueing model of LTE-U.}
	\label{vacation}
\end{figure}

In this paper, we assume that LTE-U has a listen-before-talk feature, which is likely to be a dominant mode in emerging LTE-U standards \cite{jindal2015lte}. In this mode, an AP operating LTE-U with data to send first performs carrier sensing before its transmission. We use queues with vacation and nonexhaustive service \cite{Tian:Vacation} to model such a mechanism. The queueing scheme is depicted in Fig. \ref{vacation}. Specifically, the server of queue $j$ will take a single vacation after completing the service of each packet. The vacation duration $V_j$ (in seconds) is a random variable following a certain distribution. The mean and variance of the vacation time depend on the level of local interference,\footnote{Although coexistence issues in the unlicensed band are very important over fast timescales, over slow timescales, effective interference levels and associated packet delays suffice to characterize how LTE-U interacts with other devices in the unlicensed band.} such as interference from WiFi users. The higher the interference level, the more time a queue has to wait before being served.

\section{Queueing Delays and A Conservative Allocation Scheme}
In this section, we adopt the conservative service rate model (\ref{eq4}) to develop a scheme for allocating licensed and unlicensed spectrum. For each RAT, the corresponding $k$ UE groups form $k$ independent $M/M/1$ queues, because the service rate of each queue is independent of the states of other queues under the model (\ref{eq4}). Let RAT $1$ represent LTE and RAT $2$ represent LTE-U.

Under LTE, the average packet delay (in seconds/packet) of UE group $j$ is given by the average delay for the $M/M/1$ queue:
\begin{equation}
t^{(1)}_j=\frac{1}{\left(r^{(1)}_j-\lambda^{(1)}_j\right)^+}, 
\end{equation}
where $\left(x\right)^+ = x$ if $x>0$ and $(x)^+ = 0$ if $x\leq 0$. If $r^{(1)}_j \leq \lambda^{(1)}_j$, the queueing delay is infinite, i.e., the queue becomes unstable. It is important to note that the right hand side of (7) is convex in the pair $(r^{(1)}_j, \lambda^{(1)}_j)$ on $\mathds{R}^2$. 

Under LTE-U, the $k$ UE groups form $k$ independent $M/M/1$ queues with single vacation. We have:

\textit{Proposition 1}: The average packet delay (in seconds/packet) of a queue with Poisson arrival of rate $\lambda$ and a single server of rate $r$ with single vacation $V$ whose expected squared duration is $\nu=E[V^2]$ is given by:
\begin{equation}
t = \frac{2+r \lambda\nu}{2\left(r-\lambda\right)^+}. 
\end{equation}
For fixed $\nu$, the function defined by (8) is convex in the pair $(r,\lambda)$ on $\mathds{R}^2$.

The proof of Proposition 1 is given in Appendix A.

The joint user association and conservative spectrum allocation problem is formulated as P1:
\begin{alignat*}{2}
\mathop{\text{minimize}}\limits_{\bm{r},\bm{x},\bm{y},\bm{\lambda},\bm{t}}\ \ \ \ &\frac{1}{\sum\limits_{j\in \mathcal{K}}\lambda_{j}} \sum_{j\in \mathcal{K}} (\lambda^{(1)}_jt^{(1)}_j+\lambda^{(2)}_jt^{(2)}_j)\tag{P1a}\\
\text{subject to} \ \ \ \
&t^{(1)}_j=\frac{1}{\left(r^{(1)}_j-\lambda^{(1)}_j\right)^+}, &j\in \mathcal{K} \tag{P1b}\\
&t^{(2)}_j = \frac{2+r^{(2)}_j \lambda^{(2)}_j\nu_j}{2\left(r^{(2)}_j-\lambda^{(2)}_j\right)^+},\ \ \ \ \ \ \ \ \ \  \tag{P1c}  &j\in \mathcal{K}\\
&r^{(l)}_j=\sum\limits_{\mathcal{A}\subset \mathcal{N}}\sum\limits_{i\in \mathcal{N}}s_{i \rightarrow j}^{\mathcal{A},l}x_{i \rightarrow j}^{\mathcal{A},l}, \tag{P1d}\ \ \ \ \  &j\in \mathcal{K}, l \in \{1,2\}\\
&\lambda^{(1)}_j+\lambda^{(2)}_j\geq \lambda_{j}, \tag{P1e}\ \ \ \ \ \ \ \ \ \ \ \ \ \ \ \ \ \ \ \ \ \ \ \ \ \ &j\in \mathcal{K}\\
&y^{\mathcal{A},l} \geq \sum\limits_{j\in \mathcal{K}}x_{i \rightarrow j}^{\mathcal{A},l}, \tag{P1f}\ \ \ \ \ \ &l \in \{1,2\}, i\in \mathcal{N}, \mathcal{A} \subset \mathcal{N} \\
&\sum\limits_{\mathcal{A} \subset \mathcal{N}}y^{\mathcal{A},l}=1,\ \ \ \ \ \ \ \ \ \ \ \ \ \ \ \ \ \ \ \ \ \ \ \ \ \ \ &l \in \{1,2\}\tag{P1g}\\
&x_{i \rightarrow j}^{\mathcal{A},l} \geq 0,\ \ \ \ &j\in \mathcal{K}, l \in \{1,2\}, i\in \mathcal{N}, \mathcal{A} \subset \mathcal{N}.\tag{P1h}
\end{alignat*}
The objective (P1a) is the average packet delay of all queues over the entire network. (P1b) is the average packet delay of each UE group served by LTE as derived in (7). (P1c) is the average packet delay of each UE group served by LTE-U as derived in (8). (P1d) is the service rate of each divided UE group using the conservative model as given in (4). (P1e) is the total traffic constraint for each user type (6). (P1f) guarantees the consistency of the spectrum allocation as given in (2). (P1g) constrains the total bandwidth of each RAT to be one unit. (P1h) precludes negative bandwidth. The variables in P1 are $\bm{r},\bm{x},\bm{y},\bm{\lambda},\bm{t}$ which are the vector forms of $(r^{(l)}_j)_{j\in \mathcal{K},l \in \{1,2\}}$, $(x_{i \rightarrow j}^{\mathcal{A},l})_{j\in \mathcal{K},i\in \mathcal{N},\mathcal{A} \subset \mathcal{N},l \in \{1,2\}}$, $(y^{\mathcal{A},l})_{\mathcal{A}\subset \mathcal{N},l \in \{1,2\}}$, $(\lambda^{(l)}_j)_{j\in \mathcal{K}, l \in \{1,2\}}$, $(t^{(l)}_j)_{j\in \mathcal{K}, l \in \{1,2\}}$, respectively.

P1 is a bi-convex optimization problem because when variable $\bm{\lambda}$ (resp.\ $\bm{x}$) is fixed, all constraints are linear and the objective is a linear combination of convex functions in $\bm{x}$ (resp.\ $\bm{\lambda}$). Therefore, P1 can be solved by alternating optimization over $\bm{\lambda}$ and $\bm{x}$. Since the sequence of objective values obtained at each iteration is lower-bounded and non-increasing, the objective converges although it may not be a global minimum. As shown in Section VI, the alternating method achieves good performance.

\textit{Theorem 1}: The (global) minimum average delay can be achieved by a sparse allocation, where the spectrum of each RAT is divided into at most $k$ segments. That is, there exists an optimal solution that satisfies
\begin{equation}
|\{\mathcal{A}~|~y^{\mathcal{A},l}>0,\mathcal{A}\subset \mathcal{N}\}|\leq k, \ \ \ \ \forall l \in \mathcal{M} 
\end{equation}
where $\mid\cdot\mid$ denotes the cardinality of a set. The same applies to the delay achieved by the alternating optimization method.

Theorem 1 is proved in Appendix B. The theorem guarantees that although the number of all possible patterns grows exponentially with the number of APs in the network, using a small number of patterns achieves the optimal performance.

\textit{Theorem 2}: Assume the channel gains are jointly continuous random variables. The (global) minimum average delay can be achieved with at most $n-1$ user groups served by multiple APs on each RAT. That is, there exists an optimal solution that satisfies
\begin{equation}
\left|\{j~| x_{i_{1} \rightarrow j}^{\mathcal{A}_1,l}, x_{i_{2} \rightarrow j}^{\mathcal{A}_2,l}>0,~\text{for some}~i_1,i_2\in \mathcal{N}~\text{and}~\mathcal{A}_1, \mathcal{A}_2\subset \mathcal{N}\}\right|\leq n-1, \ \ \ \ \forall l \in \mathcal{M}.  
\end{equation}
The same applies to the delay achieved by the alternating optimization method.

Theorem 2 is proved in Appendix C. The theorem guarantees that although we allow a UE group to be served by multiple APs, most UE groups will be associated with only one AP in the optimal solution.

\section{A Utilization-Dependent Allocation Scheme}
In this section, we adopt the utilization-dependent service rate model (5). Unlike in the conservative scheme, we now let each AP adapt its transmission rates to the instantaneous set of active interfering APs. Thus the service rate is in general higher than the conservative rate.

In a stable interactive queueing system, each AP transmits (over each RAT) for a fraction of the time, referred to as the {\em utilization}. Denote the utilization of AP $i$ over RAT $l$ as $\rho^{(l)}_i\in [0,1]$. The analysis of interactive queueing system is difficult. As an approximation, we assume that different APs transmit independently over each RAT. The probability of an active interfering AP set $\mathcal{I}$ over RAT $l$ is then:
\begin{equation}
p^{\mathcal{I},l} = \left(\prod_{i \in \mathcal{I}}\rho^{(l)}_i\right)\left(\prod_{i'\notin \mathcal{I}}(1-\rho^{(l)}_{i'})\right). 
\end{equation}
When UE group $j$ is served, its instantaneous service rate is one of $2^{n}$ possible values depending on the set of active APs. Thus we use a certain ``average'' of $2^n$ independent $M/M/1$ queues to approximate the queueing behavior of UE group $j$ with interactive queues. Such an approximation is reasonable under the premise that for any AP, the influence of other APs is adequately represented by their steady state probability distribution. Under each RAT, the average delay of UE group $j$ is calculated as the expected delay over $2^n$ possible rates:
\begin{alignat*}{2}
&t_{j}^{(1)}=\sum_{\mathcal{I}\subset \mathcal{N}}\frac{p^{\mathcal{I},1}}{(r^{\mathcal{I},1}_j-\lambda^{(1)}_j)^+} \tag{12a}\\
&t_{j}^{(2)} = \sum_{\mathcal{I}\subset \mathcal{N}}\frac{p^{\mathcal{I},2}(2+r^{\mathcal{I},2}_j \lambda^{(2)}_j\nu_j)}{2(r^{\mathcal{I},2}_j-\lambda^{(2)}_j)^+}. \tag{12b}
\end{alignat*}
The utilization of UE group $j$ is calculated as its expected utilization over all possible sets of interfering APs. Specifically, when the active set of APs is $\mathcal{I}$, the fraction of time that group $j$ is served is $\frac{\lambda^{(l)}_j}{r^{\mathcal{I},l}_j}$, the average utility is obtained as:
\begin{equation}
\sigma^{(l)}_j=\sum_{\mathcal{I}\subset \mathcal{N}}p^{\mathcal{I},l}\frac{\lambda^{(l)}_j}{r^{\mathcal{I},l}_j}, \ \ \ \ \ \ \ \ \ l \in \{1,2\}. \tag{13}
\end{equation}
Since AP $i$ may serve multiple UE groups over each RAT, the utilization of AP $i$ depends on the utilization of its associated UE groups. We approximate the utilization of AP $i$ over RAT $l$ as its average utilization over the spectrum used. Specifically, the average amount of spectrum used by AP $i$ to serve its UEs is $\sum\limits_{j\in \mathcal{K}}\sigma^{(l)}_j\sum\limits_{\mathcal{A}:i\in \mathcal{A}}x_{i \rightarrow j}^{\mathcal{A},l}$. Hence AP $i$'s utilization is approximated as 
\begin{equation}
\rho^{(l)}_i = \frac{1}{\sum\limits_{\mathcal{A}:i\in \mathcal{A}}y^{\mathcal{A},l}}\sum_{\mathcal{A}:i\in \mathcal{A}}\sum_{j\in \mathcal{K}}\sigma^{(l)}_jx_{i \rightarrow j}^{\mathcal{A},l},\tag{14}
\end{equation}
where $\sum\limits_{\mathcal{A}:i\in \mathcal{A}}y^{\mathcal{A},l}$ is the total bandwidth used by AP $i$ over RAT $l$.

With the above approximation, the joint user association and utilization-dependent spectrum allocation problem is formulated as P2:
\begin{alignat*}{2}
\mathop{\text{minimize}}\limits_{\bm{r},\bm{x},\bm{y},\bm{\lambda},\bm{t},\bm{\sigma},\bm{\rho},\bm{p}}\ \ \ \ &\frac{1}{\sum\limits_{j\in \mathcal{K}}\lambda_{j}} \sum_{j\in \mathcal{K}} (\lambda^{(1)}_jt_{j}^{(1)}+\lambda^{(2)}_jt_{j}^{(2)})\tag{P2a}\\
\text{subject to} \ \ \ \
&t_{j}^{(1)}=\sum_{\mathcal{I}\subset \mathcal{N}}\frac{p^{\mathcal{I},1}}{\left(r^{\mathcal{I},1}_j-\lambda^{(1)}_j\right)^+}, &j\in \mathcal{K} \tag{P2b}\\
&t_{j}^{(2)} = \sum_{\mathcal{I}\subset \mathcal{N}}\frac{p^{\mathcal{I},2}(2+r^{\mathcal{I},2}_j \lambda^{(2)}_j\nu_j)}{2\left(r^{\mathcal{I},2}_j-\lambda^{(2)}_j\right)^+},\tag{P2c}  &j\in \mathcal{K}\\
&r^{\mathcal{I},l}_j=\sum\limits_{\mathcal{A}\subset \mathcal{N}}\sum\limits_{i\in \mathcal{N}}s_{i \rightarrow j}^{\mathcal{A}\cap\mathcal{I},l}x_{i \rightarrow j}^{\mathcal{A},l}, \tag{P2d} &\mathcal{I} \subset \mathcal{N}, j\in \mathcal{K}, l \in \{1,2\}\\
&\lambda^{(1)}_j+\lambda^{(2)}_j\geq \lambda_{j}, \tag{P2e} \ \ \ \ \ \ \ \ \ \ \ \ \ \ \ \ \ \ \ &j\in \mathcal{K}\\
&y^{\mathcal{A},l} \geq \sum\limits_{j\in \mathcal{K}}x_{i \rightarrow j}^{\mathcal{A},l}, \tag{P2f} \ \ \ \ \ \ \ &l \in \{1,2\}, i\in \mathcal{N}, \mathcal{A} \subset \mathcal{N} \\
&\sum\limits_{\mathcal{A} \subset \mathcal{N}}y^{\mathcal{A},l}=1,\ \ \ \ \ \ \ \ \ \ \ \ \ \ \ \ \ \ \ \ \ \ &l \in \{1,2\}\tag{P2g}\\
&x_{i \rightarrow j}^{\mathcal{A},l} \geq 0, \ \ \ \ \ \ \ \ \ &j\in \mathcal{K}, l \in \{1,2\}, i\in \mathcal{N}, \mathcal{A} \subset \mathcal{N} \tag{P2h}\\
&p^{\mathcal{I},l} = \left(\prod_{i \in \mathcal{I}}\rho^{(l)}_i\right)\left(\prod_{i'\notin \mathcal{I}}(1-\rho^{(l)}_{i'})\right), \ \ \ \  &l \in \{1,2\}, \mathcal{I} \subset \mathcal{N} \tag{P2i}\\
&\rho^{(l)}_i = \frac{1}{\sum\limits_{\mathcal{A}:i\in \mathcal{A}}y^{\mathcal{A},l}}\sum_{\mathcal{A}:i\in \mathcal{A}}\sum_{j\in \mathcal{K}}\sigma^{(l)}_jx_{i \rightarrow j}^{\mathcal{A},l},  \tag{P2j} &l \in \{1,2\}, i\in \mathcal{N} \\
&\sigma^{(l)}_j\geq\sum_{\mathcal{I}\subset \mathcal{N}}p^{\mathcal{I},l}\frac{\lambda^{(l)}_j}{r^{\mathcal{I},l}_j}, \ \ \ \ \ \ &l \in \{1,2\}, j\in \mathcal{K}.  \tag{P2k}
\end{alignat*}
The objective (P2a) is also average packet delay of all queues of the entire network. (P2d) is the service rate of each divided UE group using utilization model as in (5). The variables in P2 are $\bm{r},\bm{x},\bm{y},\bm{\lambda},\bm{t},\bm{\sigma},\bm{\rho},\bm{p}$ which are of the vector forms of $(r^{\mathcal{I},l}_j)_{\mathcal{I}\subset \mathcal{N},j\in \mathcal{K},l \in \{1,2\}}$, $(x_{i \rightarrow j}^{\mathcal{A},l})_{j\in \mathcal{K},i\in \mathcal{N},\mathcal{A} \subset \mathcal{N},l \in \{1,2\}}$, $(y^{\mathcal{A},l})_{\mathcal{A}\subset \mathcal{N},l \in \{1,2\}}$, $(\lambda^{(l)}_j)_{j\in \mathcal{K},l \in \{1,2\}}$, $(t_{j}^{(l)})_{j\in \mathcal{K}, l \in \{1,2\}}$, $(\sigma^{(l)}_j)_{j\in \mathcal{K},l \in \{1,2\}}$, \\
$(\rho^{(l)}_i)_{i\in \mathcal{N},l \in \{1,2\}}$, and $(p^{\mathcal{I},l})_{\mathcal{I} \subset \mathcal{N}, l \in \{1,2\}}$, respectively. P2 would be equivalent to P1 if we let $p^{\mathcal{N},l}=1$, $p^{\mathcal{I},l} = 0$ for every $\mathcal{I}\neq \mathcal{N}$, and $\sigma^{(l)}_j=1$ for every $l$ and $j$. (This is a feasible suboptimal solution.) Unlike P1, P2 is not bi-convex because new variables and nonlinear constraints are introduced. However, when fixing the variables $\bm{\sigma},\bm{\rho}$ and $\bm{p}$, it becomes bi-convex. Therefore, we divide P2 into two subproblems and solve them alternately to update all the variables.

\begin{algorithm}
	\caption{Update $\bm{\sigma},\bm{\rho}$ and $\bm{p}$}
	\label{alg1}
	\begin{algorithmic}
		\STATE \textbf{Step 1. Initialization:} set initial utilization of UE groups $\bm{\sigma}(0)$;
		\STATE \textbf{Step 2. Update:} ${\sigma^{(l)}_{j}}(m) \leftarrow \sum\limits_{\mathcal{I}\subset \mathcal{N}}p^{\mathcal{I},l}(m-1)\frac{\lambda^{(l)}_j}{r^{\mathcal{I},l}_j}, \ \ \ \ \ \ \ \ \ \ \ \ \ \ \ \ \ \ \ \ \ \ \ \ \ \ \ \  l \in \{1,2\}, j\in \mathcal{K}$\\
		~~~~~~~~~~~~~~~~~~~~~${\rho^{(l)}_{i}}(m) \leftarrow \frac{1}{\sum\limits_{\mathcal{A}:i\in \mathcal{A}}y^{\mathcal{A},l}}\sum\limits_{\mathcal{A}:i\in \mathcal{A}}\sum\limits_{j\in \mathcal{K}}{\sigma^{(l)}_{j}}(m)x_{i \rightarrow j}^{\mathcal{A},l}, \ \ \ \ \ \ \ \ \ \ \ \ \ \ \ \ \ l \in \{1,2\}, i\in \mathcal{N}$\\
		~~~~~~~~~~~~~~~~~~~~~$p^{\mathcal{I},l}(m) \leftarrow \left(\prod\limits_{i \in \mathcal{I}}{\rho^{(l)}_{i}}(m)\right)\left(\prod\limits_{i'\notin \mathcal{I}}(1-{\rho^{(l)}_{i'}}(m))\right),  \ \ \ \ \ \ \ \ \ \  l \in \{1,2\}, \mathcal{I} \subset \mathcal{N}$\\
		~~~~~~~~~~~~~~~~~~~~~$m\leftarrow m+1;$
		\STATE \textbf{Step 3.} if $\parallel \bm{\sigma}(m)-\bm{\sigma}(m-1)\parallel\geq \epsilon$, where $\epsilon$ is a small number, go back to Step 2; otherwise stop and obtain\\
		$\sigma^{(l)}_j \leftarrow {\sigma^{(l)}_{j}}(m)$, $\rho^{(l)}_i \leftarrow {\rho^{(l)}_{i}}(m)$, $p^{\mathcal{I},l} \leftarrow p^{\mathcal{I},l}(m)$, ~~~~~~~~~~~~~ $l \in \{1,2\}, j\in \mathcal{K}, i\in \mathcal{N}, \mathcal{I} \subset \mathcal{N}$.
	\end{algorithmic}
\end{algorithm}

\begin{algorithm}
	\caption{Iterative algorithm for solving P2}
	\label{alg2}
	\begin{algorithmic}
		\STATE \textbf{Initialization:} $\bm{x} \leftarrow 0; \bm{x}' \leftarrow 1; \sigma^{(l)}_j \leftarrow 1, \forall j\in \mathcal{K},l \in \{1,2\}; \rho^{(l)}_i \leftarrow 1, \forall i\in \mathcal{N}, l \in \{1,2\}; p^{\mathcal{N},l} \leftarrow 1$, $p^{\mathcal{I},l} \leftarrow 0$, $\forall \mathcal{I}\subset \mathcal{N}$, $\mathcal{I} \neq \mathcal{N}$.
		\WHILE {$|| \bm{x}-\bm{x}'|| > \epsilon$}
		\STATE {1. $\bm{x}' \leftarrow \bm{x}$;}
		\STATE {2. Update $\bm{r},\bm{x},\bm{y},\bm{\lambda},\bm{t}$ by solving P3;}
		\STATE {3. Update $\bm{\sigma},\bm{\rho},\bm{p}$ by using Algorithm 1.}
		\ENDWHILE
	\end{algorithmic}
\end{algorithm}

Given $\bm{\sigma},\bm{\rho}$ and $\bm{p}$, we update $\bm{r},\bm{x},\bm{y},\bm{\lambda},\bm{t}$ by solving subproblem P3:
\begin{alignat*}{2}
\mathop{\text{minimize}}\limits_{\bm{r},\bm{x},\bm{y},\bm{\lambda},\bm{t}}\ \ \ \ &\frac{1}{\sum\limits_{j\in \mathcal{K}}\lambda_{j}} \sum_{j\in \mathcal{K}} (\lambda^{(1)}_jt_{j}^{(1)}+\lambda^{(2)}_jt_{j}^{(2)})\tag{P3a}\\
\text{subject to} \ \ \ \
&\sum_{\mathcal{I}\subset \mathcal{N}}p^{\mathcal{I},l}\frac{\lambda^{(l)}_j}{r^{\mathcal{I},l}_j}\leq\sigma^{(l)}_j, \ \ \ \ \ \ &l \in \{1,2\};j\in \mathcal{K}  \tag{P3b}\\
&(\textrm{P2b})-(\textrm{P2h}).
\end{alignat*}
(P3b) constraints the utilization of UE groups. The structure of subproblem P3 is similar to that of P1. It can be solve by alternating optimization over $\bm{\lambda}$ and $\bm{x}$.

Given $\bm{r},\bm{x},\bm{y},\bm{\lambda},\bm{t}$, we update $\bm{\sigma},\bm{\rho}$ and $\bm{p}$. There are $2(2^{n}+n+k)$ variables and equations. To solve them with low complexity, Algorithm 1 updates $\bm{\sigma},\bm{\rho}$ and $\bm{p}$ iteratively as in \cite{teng2015resource,teng2015Dual}. Here $m$ denotes the iteration. Convergence of the algorithm can be established similarly as in \cite{teng2015resource,teng2015Dual}. In addition, the solution obtained by Algorithm 1 is feasible for (P2i)-(P2k).

After solving two subproblems alternately, P2 can be solved using Algorithm 2. Each step in Algorithm 2 is an easier problem. Convergence of Algorithm 2 can also be similarly established as in \cite{teng2015resource,teng2015Dual}. Therefore, P2 can be solved iteratively with low complexity.

\section{Extension}
So far, the packet lengths are assumed to be i.i.d with exponential distribution, which results in exponentially distributed service time for each packet. In this section, we show that the proposed schemes and algorithms are also applicable to general packet length distributions as long as the first and second moments if the service time can be written in the form:
\begin{alignat*}{3}
&E[X] = \frac{\beta}{r},  \tag{15a}\\
&E[X^2] = \frac{\eta}{r^2}, \tag{15b}
\end{alignat*}
where $r$ is the service rate and $\beta$ and $\eta$ are positive numbers. A wide class of distributions have such characteristics. For example, $\beta = 1$ and $\eta = 2$ correspond to exponential service time; $\beta = 1$ and $\eta = 1$ correspond to constant service time. In general, the queues in the system are not $M/M/1$ queues. In contrast to (7) and (8), the formulas for calculating the average packet delay of UE group $j$ become:
\begin{alignat*}{4}
&\hat{t}^{(1)}_j = \frac{(\frac{1}{2}\eta-\beta^2)\lambda^{(1)}_j+\beta r^{(1)}_j}  {r^{(1)}_j\left(r^{(1)}_j-\beta \lambda^{(1)}_j\right)^+}, \tag{16a}\\
&\hat{t}^{(2)}_j = \frac{(\frac{1}{2}\eta-\beta^2)\lambda^{(2)}_j+\beta r^{(2)}_j+\frac{1}{2}\nu_j \lambda^{(2)}_j{\left(r^{(2)}_j\right)}^2}  {r^{(2)}_j\left(r^{(2)}_j-\beta \lambda^{(2)}_j\right)^+}. \tag{16b}
\end{alignat*}
Then the objective function in P1 is rewritten as:
\begin{alignat*}{4}
\hat{U}= \sum_{j\in \mathcal{K}} (\lambda^{(1)}_j\hat{t}^{(1)}_j+\lambda^{(2)}_j\hat{t}^{(2)}_j).\tag{17}
\end{alignat*}
\textit{Proposition 2}: The new objective function (17) is bi-convex in $\bm{\lambda}$ and $\bm{r}$.

Proposition 2 is proved in Appendix D. Due to Proposition 2, the techniques proposed in Section III and Section IV still apply to the problem of user association and spectrum allocation with general packet length distribution .
\section{Numerical Results}

\begin{figure}[t]
\centering
\includegraphics[width=5in]{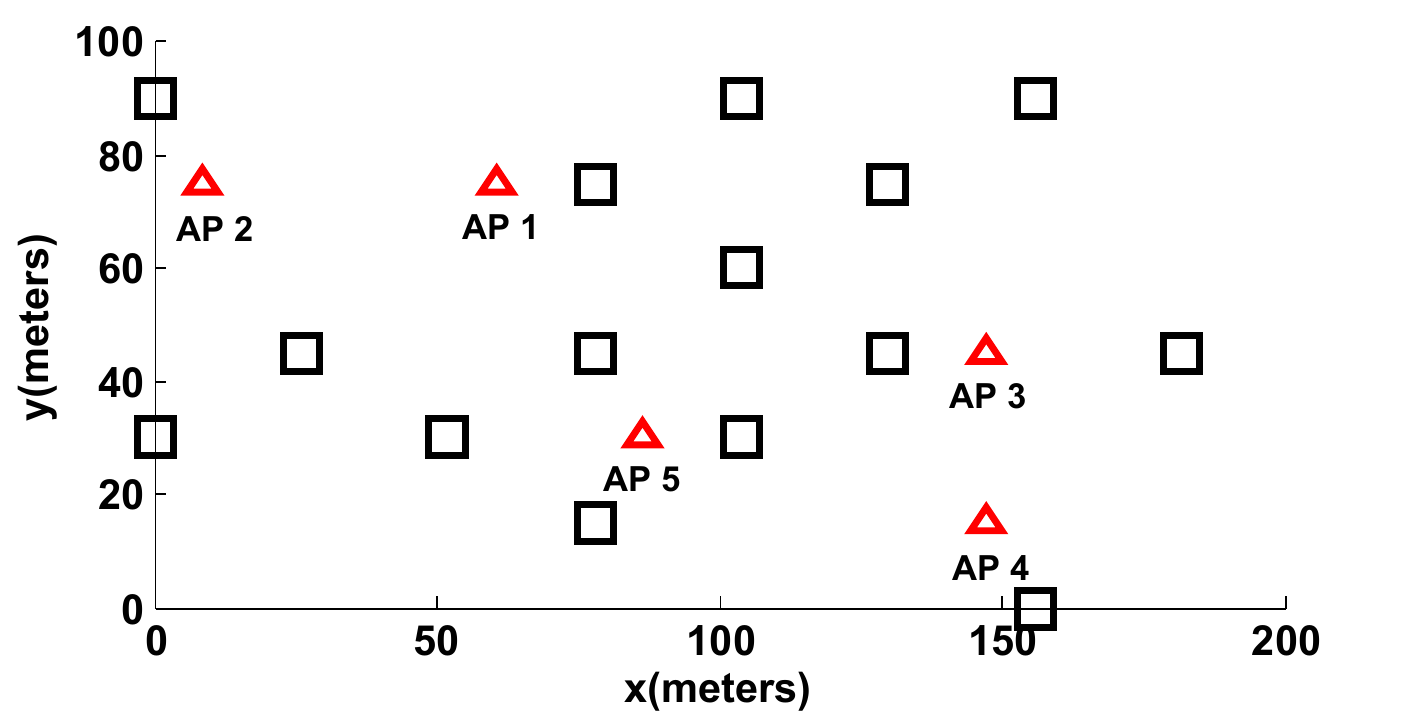}
\caption{The topology of the 5-AP network.}
\label{deployment}
\end{figure}
\begin{table}[t]
	%% increase table row spacing, adjust to taste
	\renewcommand{\arraystretch}{1.3}
	% if using array.sty, it might be a good idea to tweak the value of
	% \extrarowheight as needed to properly center the text within the cells
	\caption{Parameter Values.}
	\label{table_example}
	\centering
	\begin{tabular}{|c|c|c|}
		\hline
		Parameters & Value/Function \\
		\hline
		AP transmit power  & 23 dBm\\
		Total bandwidth on each RAT &  10 MHz\\
		Average packet length & 0.5 Mbits\\
		AP to UE pathloss & $140.7+36.7\log_{10}(R)$\\
		\hline
	\end{tabular}
\end{table}
Simulations are carried out using the network topology depicted in Fig. \ref{deployment}. The HetNet is deployed over a $100 \times 200$ $\text{m}^2$ area. Five APs, denoted by triangles, and 15 UE groups, denoted by the squares, are randomly dropped in the area. The spectral efficiency is calculated using (3) assuming $\alpha^{(1)} = 1$ and $\alpha^{(2)}=1/2$ with a 30 dB cap on the receive SINR (SINR greater than 30 dB is regarded as 30 dB). Other parameters used in the simulation are given in Table I, and are compliant with the LTE standard\footnote{Note that the pathloss models for different RATs need not be the same.} \cite{3GPP:LTE}. The results for actual packet delay with the utilization service rate model are obtained through a packet-level simulator, which adapts the transmission time of each packet to the instantaneous active interfering APs.

\begin{figure}[t]
  \centering
  \subfigure[Spectrum allocation pattern for LTE]{
    \label{fig:subfig:a} %% label for first subfigure
    \includegraphics[width=5in]{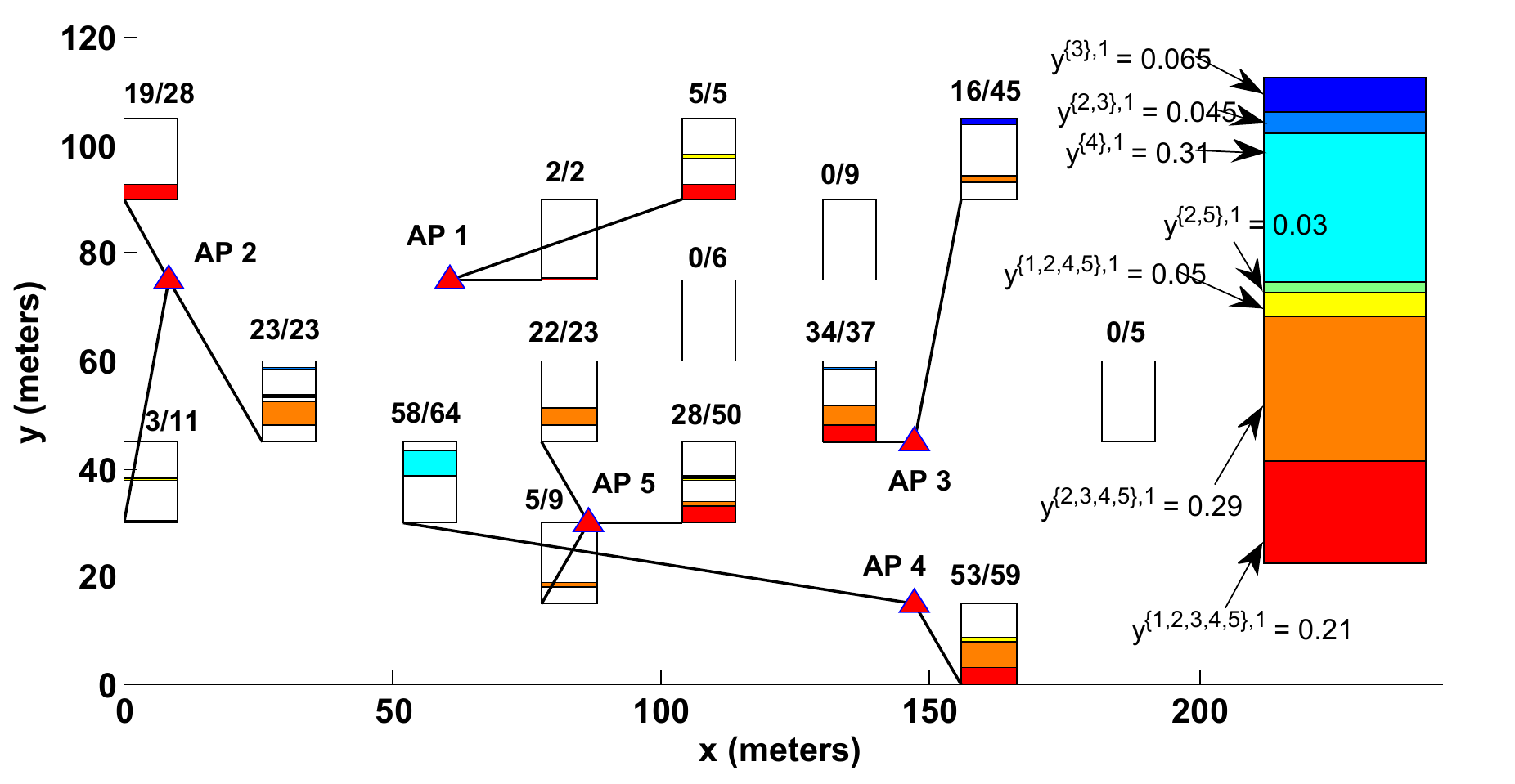}}
  \hspace{1in}
  \subfigure[Spectrum allocation pattern for LTE-U]{
    \label{fig:subfig:b} %% label for second subfigure
    \includegraphics[width=5in]{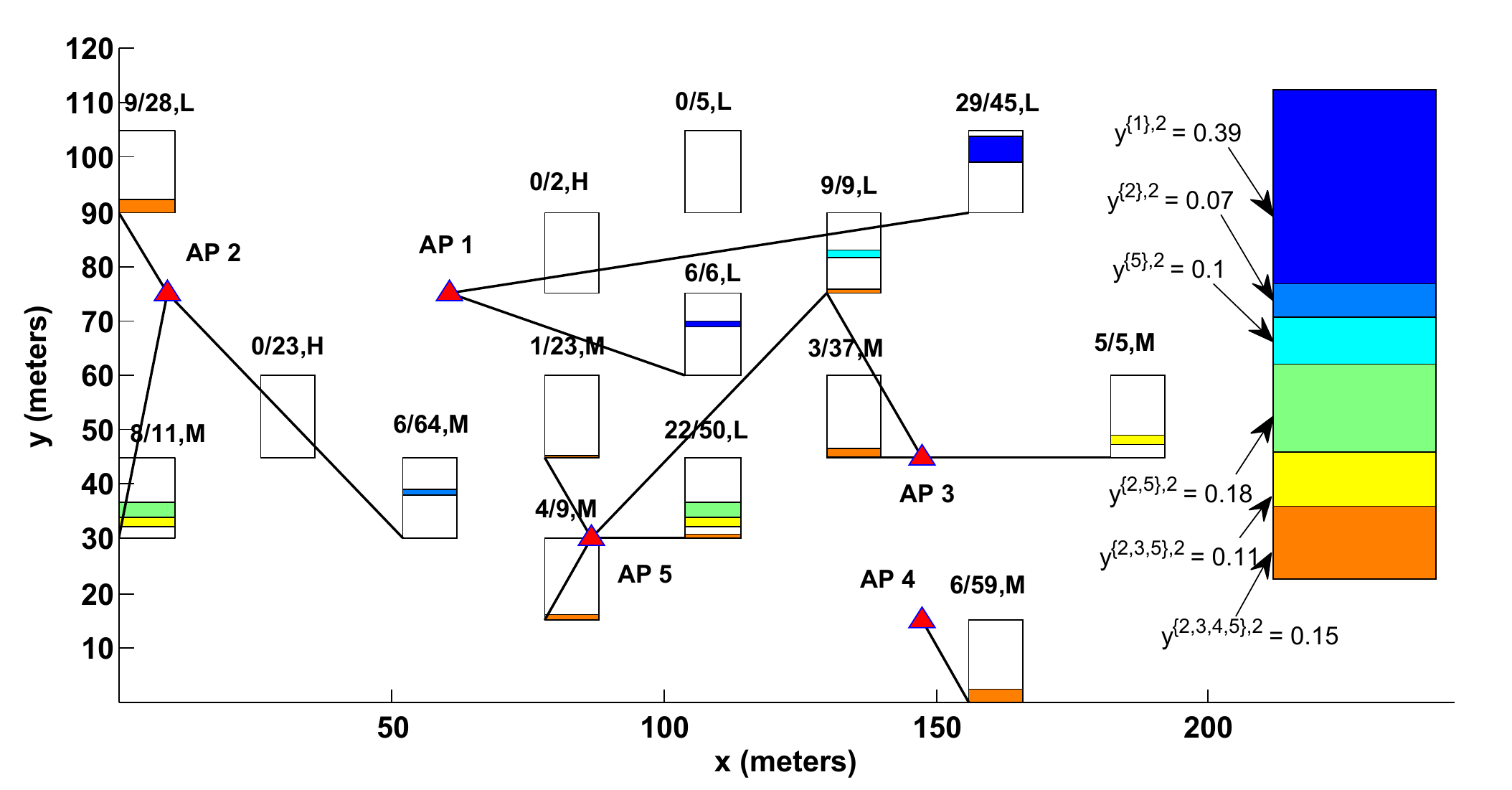}}
  \caption{Spectrum allocation patterns under heavy traffic, $\nu_{H} = 1, \nu_{M} = 0.01, \nu_{L} = 0.0025$.}
  \label{fig:subfig} %% label for entire figure
\end{figure}

\begin{figure}[t]
	\centering
	\subfigure[Spectrum allocation pattern for LTE]{
		\label{fig:subfig:a} %% label for first subfigure
		\includegraphics[width=5in]{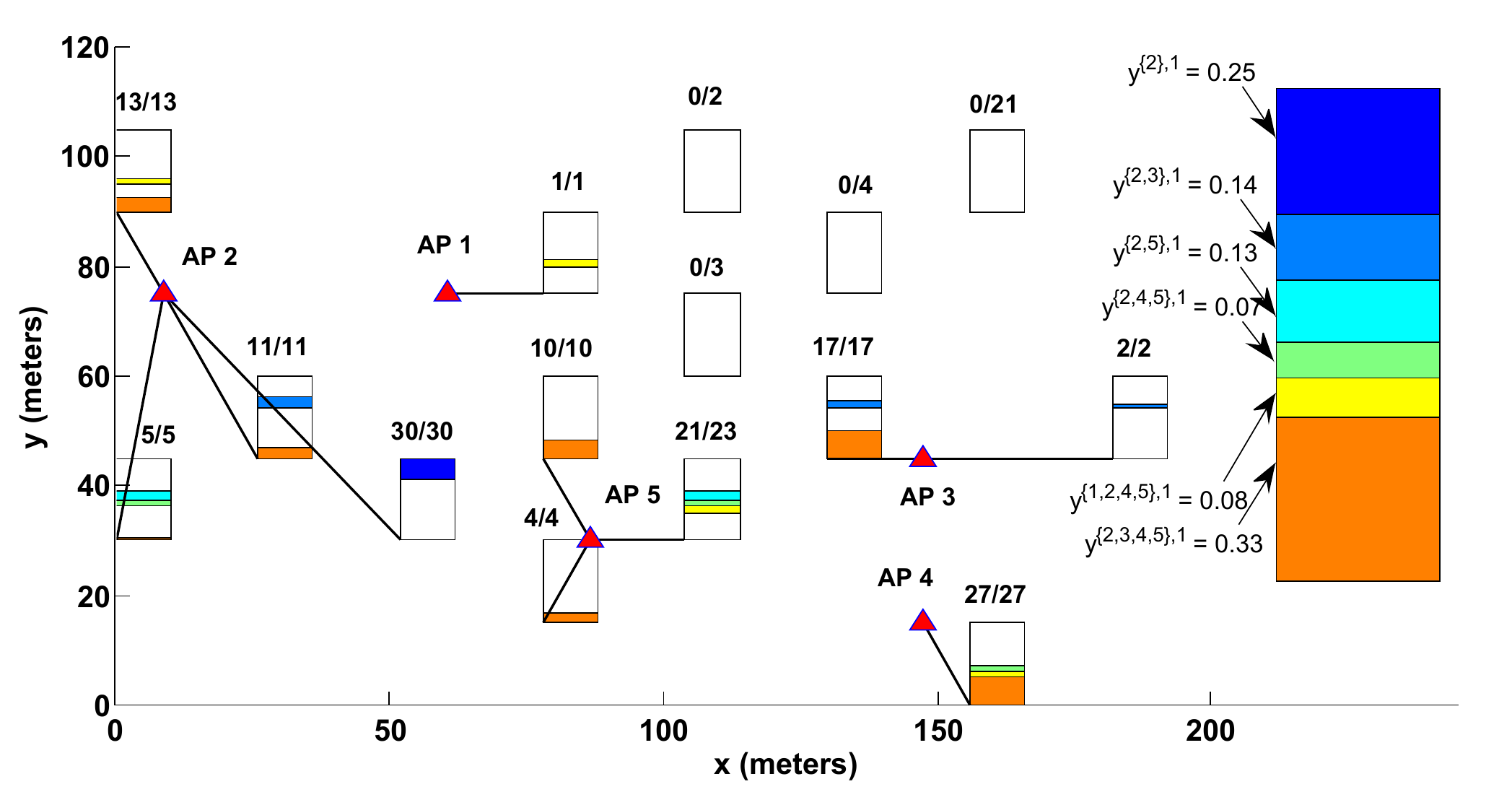}}
	\hspace{1in}
	\subfigure[Spectrum allocation pattern for LTE-U]{
		\label{fig:subfig:b} %% label for second subfigure
		\includegraphics[width=5in]{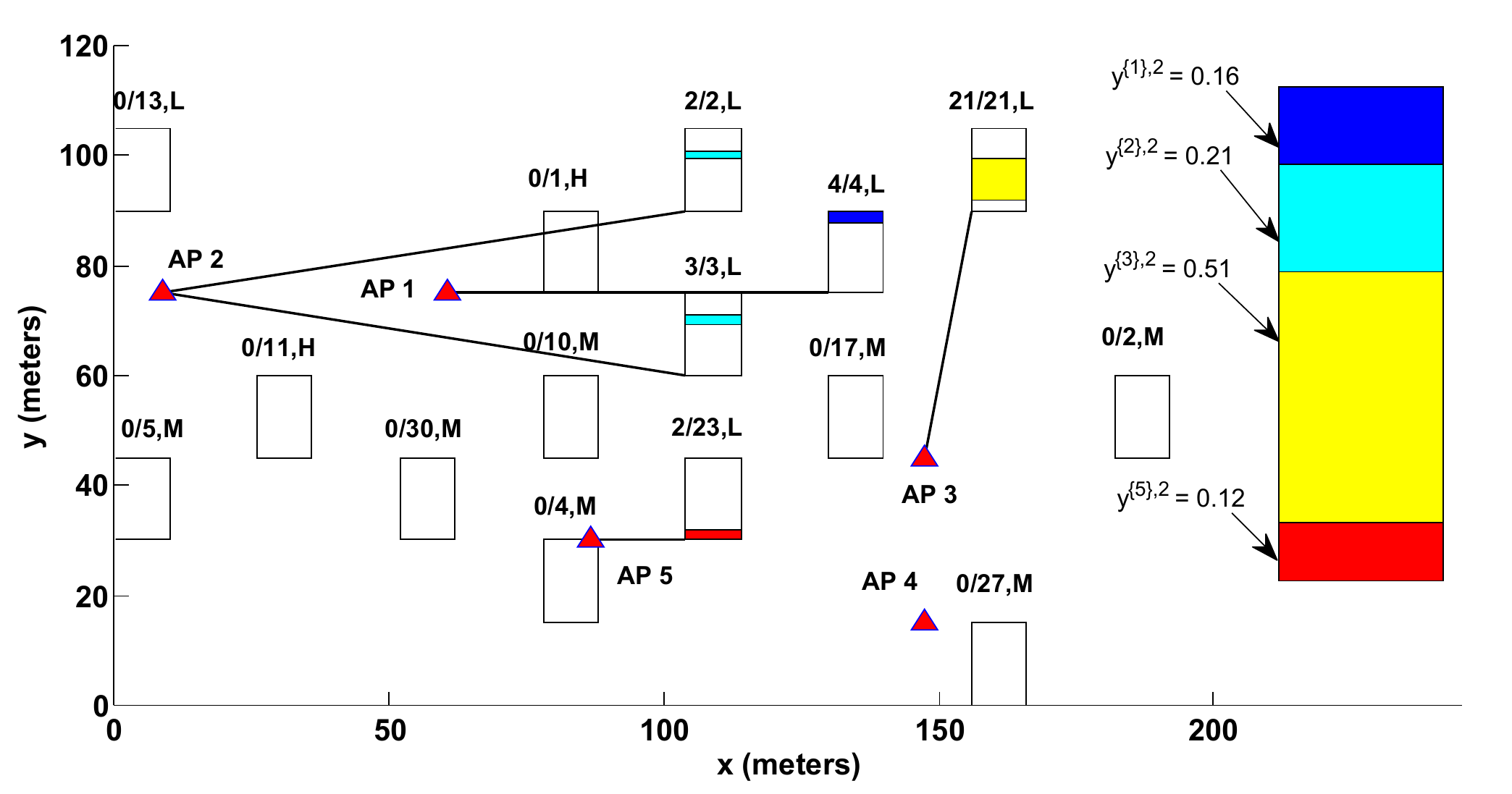}}
	\caption{Spectrum allocation patterns under light/medium traffic, $\nu_{H} = 1, \nu_{M} = 0.01, \nu_{L} = 0.0025$.}
	\label{fig:subfig} %% label for entire figure
\end{figure}

\subsection{The Conservative Allocation Scheme}
The conservative spectrum allocation and UE-AP association according to the solution of P1 under different traffic loads are shown in Figs. 4 and 5. Each smaller rectangle indicates the spectrum allocation at the corresponding UE group. The different colors represent different patterns. The amount of spectrum resources allocated to each user group under each pattern is denoted by the size of the corresponding bar inside the rectangle. The normalized average packet arrival rate relative to the total (sum) average arrival rate of each user group as well as the intensity of interference nearby from the unlicensed band are shown above the rectangle. For example, for the UE group on the top left, $19/28$ of its traffic load is assigned to be served using LTE while the remaining $9/28$ is to be served using LTE-U. There are three interference levels, referred to as ``High'', ``Medium'' and ``Low'', according to the intensity of the interference nearby from the unlicensed band. For concreteness, we let the corresponding second moments of the vacation duration in (8) be $\nu_{H} = 1, \nu_{M} = 0.01, \nu_{L} = 0.0025$. When the second moment $\nu_{L} = 0.0025$, the average vacation time is similar to the average packet service time, which is about 0.05 seconds/packet. Each line segment joining an AP and a user group describes their association. The color bars on the right of each figure shows the actual spectrum partition into different patterns. Theorem 1 and Theorem 2 can be verified by counting the number of pieces in the partition and the number of rectangles connecting multiple triangles, respectively. For instance, in Fig. 4, the licensed band is partitioned into 7 segments and the unlicensed band is partitioned into 6 segments, both numbers of partitions are smaller than the number of UE groups. Moreover, there is only one UE group served by multiple APs in the unlicensed band.

Fig. 4 shows the spectrum allocation patterns for the licensed band and unlicensed band in a high traffic scenario. The licensed band is partitioned into 7 segments and the unlicensed band is partitioned into 6 segments. As we can see, the spectrum resources allocated to each user group is roughly proportional to the corresponding traffic demand. Since the licensed band alone cannot support all the UE groups, some UE groups with low traffic demand are served by using the unlicensed band only and most licensed spectrum is allocated to the UE groups with high traffic. In addition, even though some UE groups have high traffic demand (e.g., the top right one), more unlicensed spectrum is allocated to them because they do not have much nearby interference from other RATs in the unlicensed band. In contrast, some UE groups with low traffic demand (e.g., the one in the middle with arrival rate 2 packets/second) may also be allocated licensed spectrum if the nearby interference from the unlicensed band is severe.

\begin{figure}[t]
	\centering
	\includegraphics[width=5in]{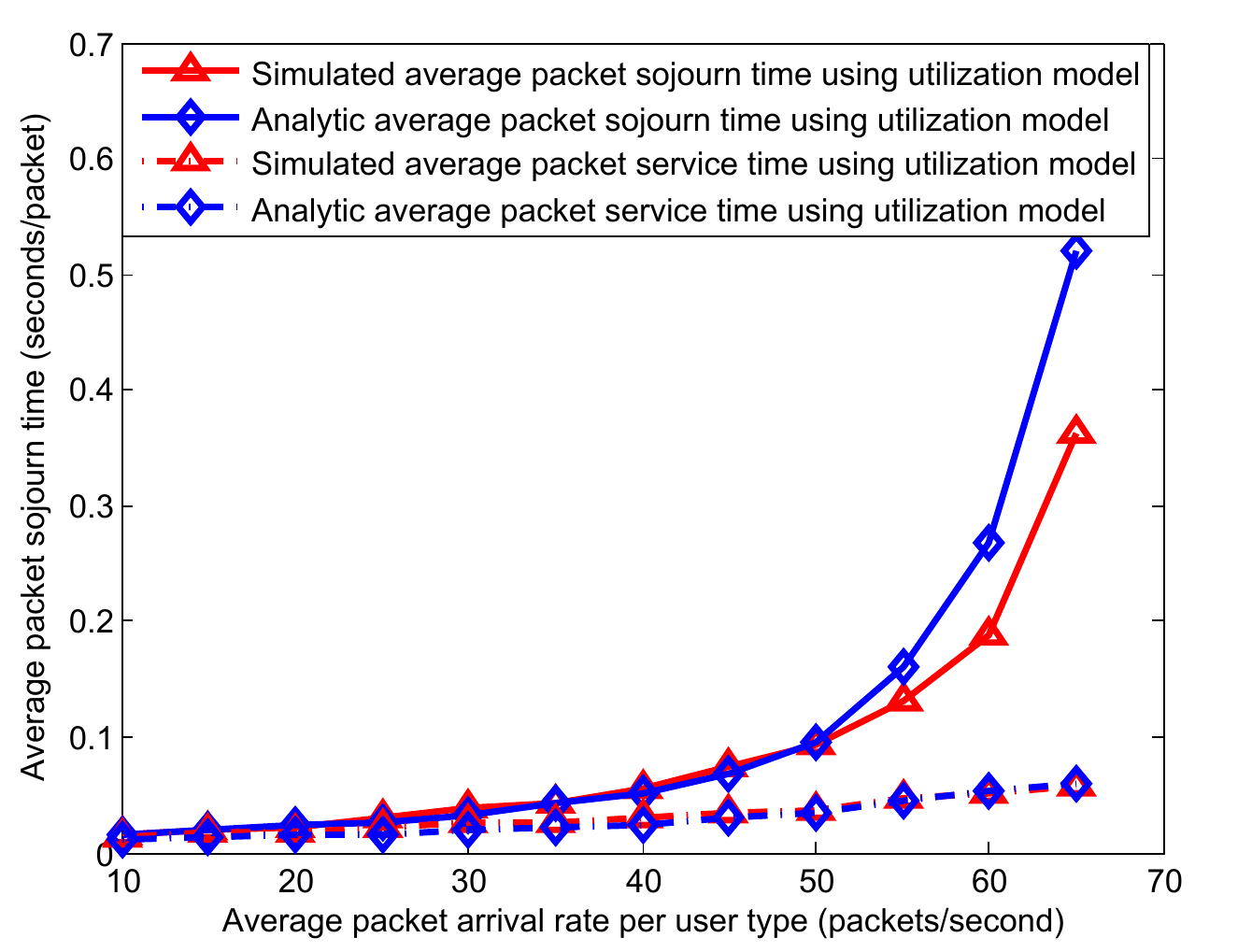}
	\caption{Analytic and simulated delay for the utilization model.}
	\label{compare}
\end{figure}

Fig. 5 shows the spectrum allocation patterns for the licensed and unlicensed bands in a medium/low traffic scenario. The licensed band is partitioned into 6 segments and the unlicensed band is partitioned into 4 segments. Since the licensed band alone is almost enough to support all the UE groups, most of them are served by the licensed band, and only a few of the UE groups are served by the unlicensed band because of the low traffic demand and lower interference generated within that band. In addition, most spectrum is allocated to the UE groups with high traffic.

\subsection{The Utilization-Dependent Allocation Scheme}

\begin{figure}[t]
\centering
\includegraphics[width=5in]{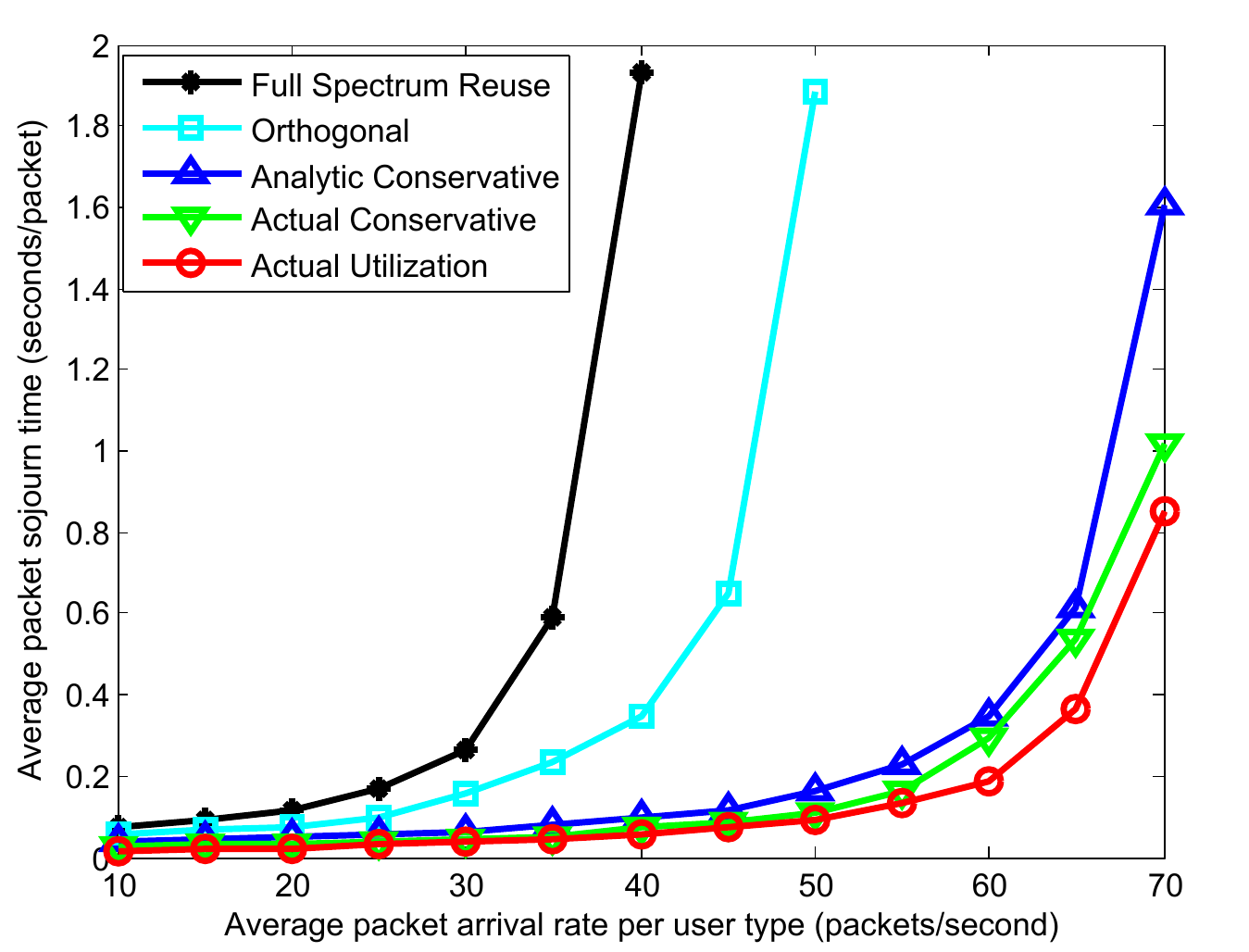}
\caption{Comparison of the proposed schemes with benchmark schemes.}
\label{compare}
\end{figure}

\iffalse
\begin{figure}[t]
  \centering
  \subfigure[Spectrum allocation pattern for LTE]{
    \label{fig:subfig:a} %% label for first subfigure
    \begin{minipage}[t]{0.5\textwidth}
    \centering
    \includegraphics[width=3.5in]{license_con_uti}
    \end{minipage}}

  \subfigure[Spectrum allocation pattern for LTE-U]{
    \label{fig:subfig:b} %% label for second subfigure
    \begin{minipage}[t]{0.5\textwidth}
    \centering
    \includegraphics[width=3.5in]{unlicense_con_uti}
    \end{minipage}}
  \caption{Optimal spectrum allocations under under different traffic loads.}
  \label{fig:subfig} %% label for entire figure
\end{figure}

\begin{figure*}
\begin{minipage}[t]{0.5\textwidth}
\centering
\includegraphics[width=3.5in]{license_con_uti}
\caption{Spectrum allocation pattern for LTE}
\label{fig:side:a}
\end{minipage}%
\begin{minipage}[t]{0.5\textwidth}
\centering
\includegraphics[width=2.95in]{unlicense_con_uti}
\caption{Spectrum allocation pattern for LTE-U}
\label{fig:side:b}
\end{minipage}
\end{figure*}
\fi

\begin{figure} \centering
\subfigure[Spectrum allocation patterns for LTE.] { \label{fig:a}
\includegraphics[width=5in]{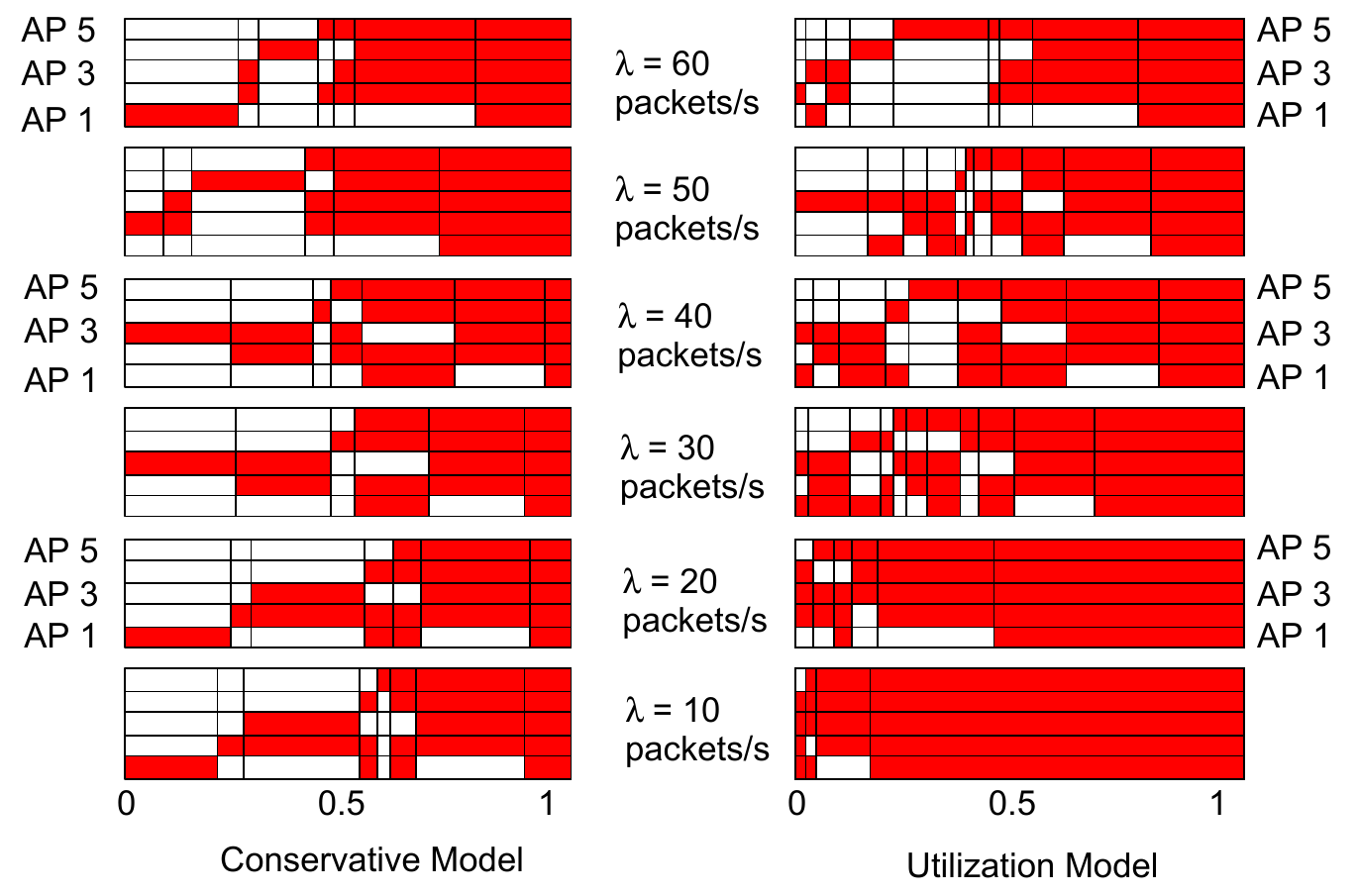}
}
\hspace{1in}
\subfigure[Spectrum allocation patterns for LTE-U.] { \label{fig:b}
\includegraphics[width=4.8in]{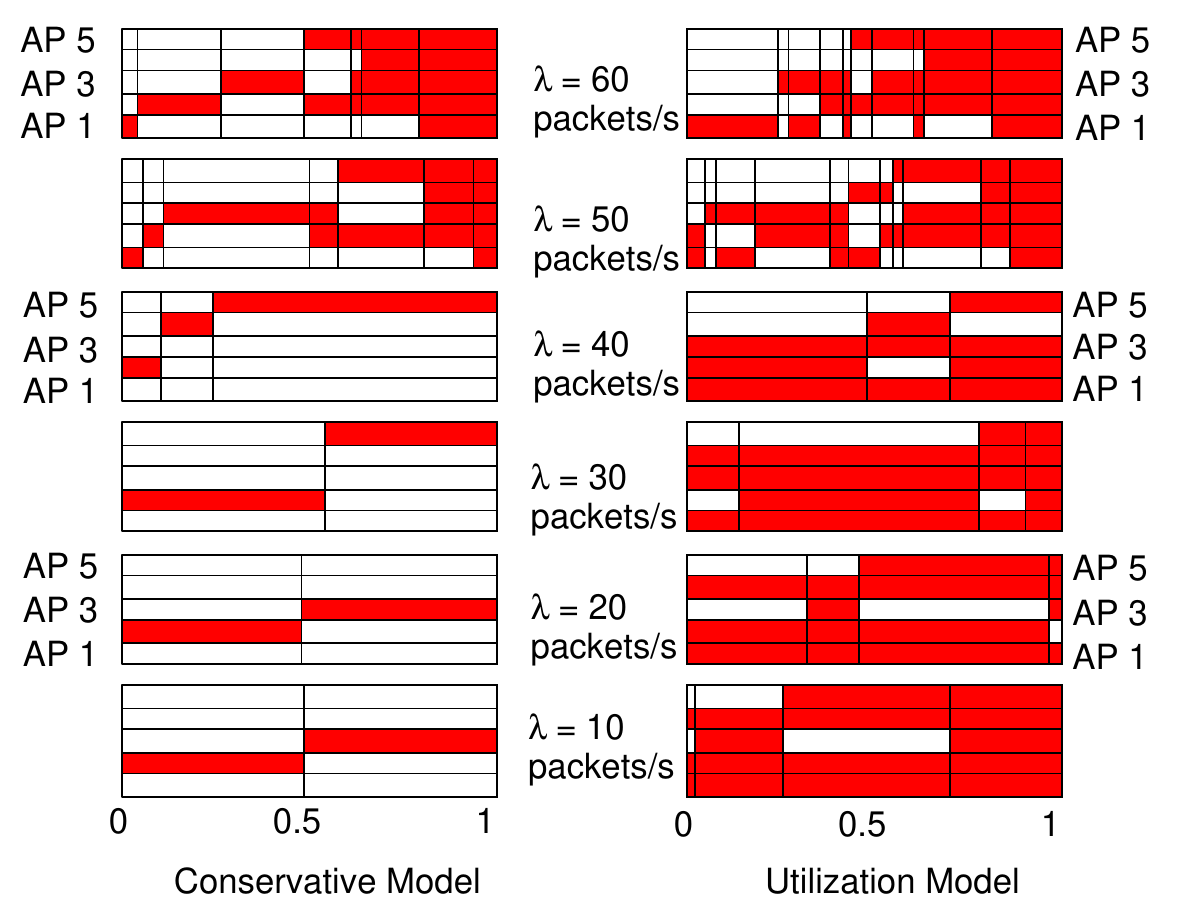}
}
\caption{Spectrum allocation patterns for different loads. }
\label{fig}
\end{figure}
Clearly, the delay obtained by using the conservative allocation is an upper bound on the actual delay obtained based on the utilization-dependent model. To validate the utilization-dependent model, a packet-level simulator is used to compare the theoretical delay with the actual delay. Fig. 6 compares theoretical and actual delays and service times using the utilization-dependent model. The service transmission time of a packet is part of the packet sojourn time excluding the queueing delay. The horizontal axis is the average traffic load of each UE group. All curves in Fig. 6 are based on the same spectrum allocation. The utilization approximation is quite accurate with light traffic. When traffic loads are heavy, the approximation becomes coarser due to more interactions among APs. In addition, compared with delay, the service time increases much more slowly with traffic load, indicating that the spectrum is efficiently allocated to mitigate interference among APs.

Fig. 7 compares the optimization objective, average packet sojourn time, for the conservative and the utilization-dependent allocation schemes with orthogonal and full-reuse allocations. As the traffic increases to 40 and 50 packets/s, respectively, full-reuse and orthogonal allocations fail to support all the UE groups, suggesting these are the maximum throughputs under these two allocation schemes, respectively. Hence, the proposed conservative and utilization-dependent allocation schemes achieve a significantly larger throughput region than the other schemes. In addition, both conservative and the utilization-dependent allocation schemes yield significant gain especially in the high traffic scenario. The reason is that the proposed spectrum allocation schemes are traffic-aware, i.e., in the low traffic regime, it is better to reuse most of the spectrum, while in the high traffic regime, it is better to adaptively allocate the spectrum across APs. Furthermore, the proposed schemes exploit the particular characteristics of the RATs. That is, UE groups receiving less external interference from the unlicensed band are more likely to be allocated spectrum from that band. Furthermore, the actual delay based on the conservative allocation is also given, which is between the theoretical delay using conservative allocation and the actual delay using the utilization-dependent allocation scheme. The utilization-dependent allocation scheme always outperforms the conservative allocation scheme as it accurately accounts for dynamic inter-cell interference. In low traffic, the allocation based on the utilization-dependent model reduces the average delay by about 40\% compared to the conservative allocation, and this improvement becomes smaller as traffic loads increase.

Fig. 8 shows the optimal allocations for the licensed and unlicensed bands under different traffic loads using the conservative and utilization-dependent allocation schemes. Rectangles represent frequency bands and solid ones are used by the corresponding APs. A stack of solid rectangles then represents a reuse pattern. The length of each rectangle is proportional to the fraction of the whole spectrum occupied by the corresponding pattern. In both Fig. 8a and Fig. 8b, when traffic loads are low, the allocation scheme using the utilization-dependent model tends to be full reuse, while as traffic loads increase, it tends to orthogonalize the spectrum. In comparison with the conservative allocation, which remains orthogonal under different traffic loads, the utilization-dependent allocation uses the spectrum more efficiently since it approximates interaction among APs more accurately.

\section{Conclusion}
We have studied spectrum allocation in downlink HetNets with multiple RATs over different bands using the average packet sojourn time as the performance metric. In addition to the licensed band, a queueing model with vacation has been proposed to model the additional delay of the unlicensed band. Two optimization based schemes have been proposed and shown to be highly effective. A sparse allocation (in terms of both the number of spectrum segments and user association) achieves the optimal network utility. Simulation results show that the proposed allocation schemes yield user associations and spectrum allocations, which utilize the spectrum of each RAT more efficiently compared with the benchmark allocation schemes, namely, orthogonal frequency reuse and full-frequency-reuse. Ongoing work is to extend the optimization framework to give a resource allocation solution with complexity that scales with the size of the network.



\begin{appendices}
	%%%Appendix A
	\section{Proof of Proposition 1} 
	To analyze the queueing model with vacation and nonexhaustive service, we follow a similar technique as in \cite{Tian:Vacation}. Denote the residual service time at time $t$ by $R(t)$, and the waiting and service times for the $i$-th packet as $W_i$ and $X_i$, respectively. We have,
	\begin{align}
	E[X] = \frac1r. \tag{17}
	\end{align} Assuming the queue is stable, by the Pollaczek-Khinchine formula,
	\begin{equation}
	E[W]=\frac{E[R]}{1-\rho}, \tag{18}
	\end{equation}
	and the average packet delay is given by:
	\begin{equation}
	E[T]=E[W]+E[X]=\frac{E[R]}{1-\rho}+\frac{1}{r}, \tag{19}
	\end{equation}
	where $\rho=\frac{\lambda}{r}$. 
	Denote $M(t)$ as the number of packets served up until time $t$. To calculate $E[R]$, we can see from Fig. 9,
	
			\begin{figure}[t]
				\centering
				\includegraphics[width=5in]{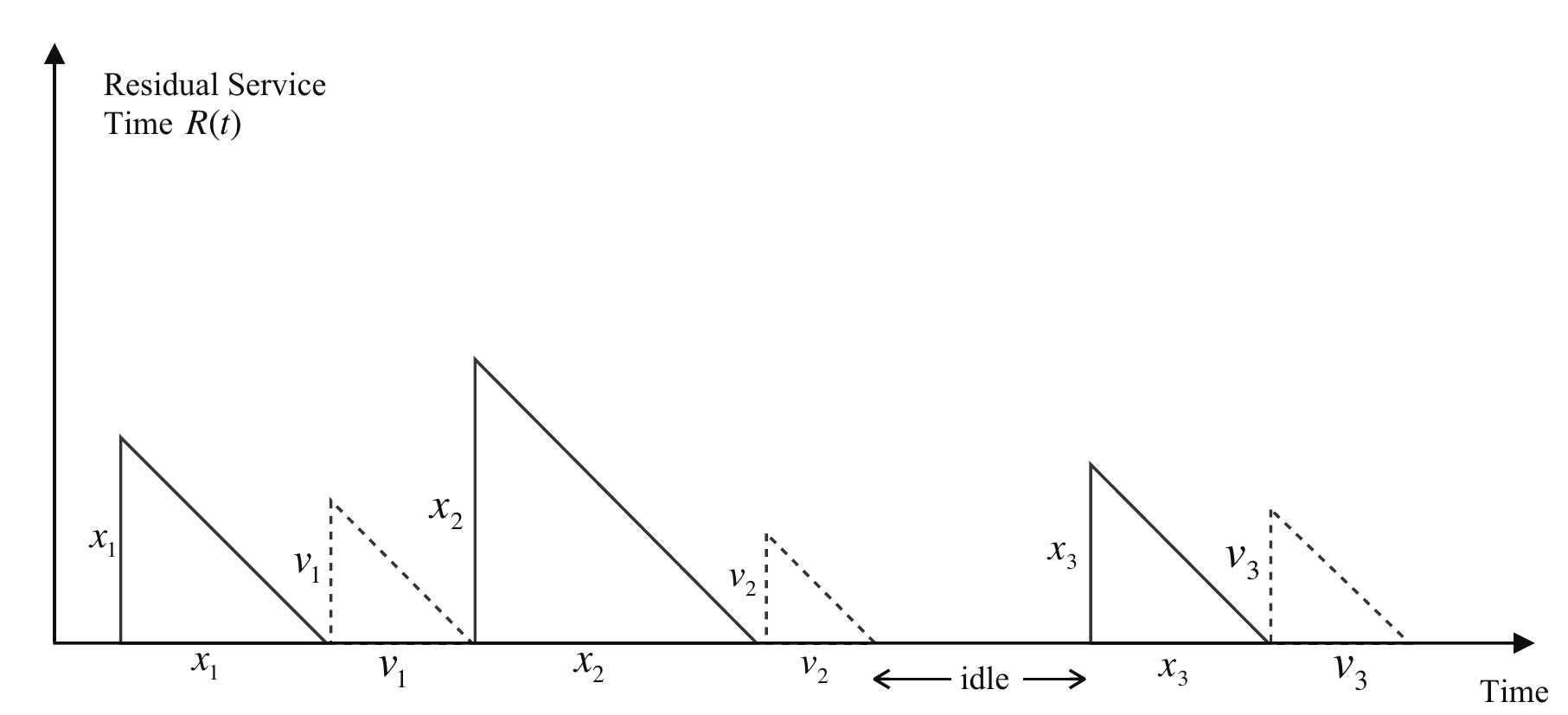}
				\caption{Residual service time \textit{R(t)}.}
				\label{hybrid}
			\end{figure}
	
	\begin{alignat*}{2}
	E[R]&=\lim_{t\rightarrow \infty} \frac{1}{t}\int^{t}_{0}R(\tau)d\tau\tag{20}\\
	&=\lim_{t\rightarrow \infty}\frac{1}{t}\sum^{M(t)}_{i=1}\left(\frac{\left(X_{i}\right)^{2}}{2}+\frac{\left(V_{i}\right)^{2}}{2}\right)\tag{21}\\
	&=\lim_{t\rightarrow \infty}\frac{M(t)}{2t}\sum^{M(t)}_{i=1}\frac{\left(X_{i}\right)^{2}+\left(V_{i}\right)^{2}}{M(t)}\tag{22}\\
	&=\frac{\lambda}{2}\left(E[X^{2}]+E[V^{2}]\right)\tag{23}\\
	&=\frac{\lambda}{r^{2}}+\frac{\lambda \nu}{2}.\tag{24}
	\end{alignat*}
	The equalities hold in the almost sure sense. Substituting (24) into (19), we obtain (7). \\
	To prove the convexity of (7), note that (7) can be written as the addition of two parts,
	\begin{equation}
	t = \frac{1}{\left(r-\lambda\right)^+} +\frac{\nu}{2}\frac{r\lambda}{\left(r-\lambda\right)^+}.  \tag{25}
	\end{equation}
	Since both parts are convex in the pair $(r,\lambda)$ on $\mathds{R}^2$, the sum is also convex in the pair $(r,\lambda)$ on $\mathds{R}^2$. \hfill $\blacksquare$

	%%%Appendix B
	\section{Proof of Theorem 1} 
	We first reformulate P1 by a change of variables. The proof then follows a similar geometric argument as in \cite{Traffic:Zhuang}.
	
	Consider a reformulation of P1 by replacing constraints (P1c), (P1e), and (P1g) with the following three constraints, introducing a new collection of variables $(z_{i \rightarrow j}^{\mathcal{A},l})$:
	\begin{alignat*}{2}
	&r^{(l)}_j=\sum_{\mathcal{A} \subset \mathcal{N}}\left(\sum_{i \in \mathcal{N}}s_{i \rightarrow j}^{\mathcal{A},l}z_{i \rightarrow j}^{\mathcal{A},l}\right)y^{\mathcal{A},l}, \tag{P4a}\ \ \  j\in \mathcal{K}, l \in \{1,2\}\\
	&\sum_{j \in \mathcal{K}}z_{i \rightarrow j}^{\mathcal{A},l} \leq 1, \tag{P4b}\ \ \ \ \ \ \ \ \ \ \ \ \ \ \ \ \ l \in \{1,2\}, i\in \mathcal{N}, \mathcal{A} \subset \mathcal{N} \\
	&z_{i \rightarrow j}^{\mathcal{A},l}, y^{\mathcal{A},l} \geq 0, \tag{P4c}\ \ \ \ \ \ \ \ j\in \mathcal{K}, l \in \{1,2\}, i\in \mathcal{N},  \mathcal{A} \subset \mathcal{N}.
	\end{alignat*}
	The new problem, referred to as P4, is equivalent to P1. This is because the feasible set for the rate tuple $\bm{r}$ remains the same, which is easy to see by regarding $z_{i \rightarrow j}^{\mathcal{A},l}$ as the fraction of spectrum under pattern $\mathcal{A}$ that AP $i$ allocates to UE group $j$ over RAT $l$. If one solves P4, the actual spectrum allocations can be recovered as $x_{i \rightarrow j}^{\mathcal{A},l}=y^{\mathcal{A},l}z_{i \rightarrow j}^{\mathcal{A},l}$.
	
	We show that if $\bm{r}^*$ is an optimal rate tuple of P1, then we can attain each sub-tuple ${\bm{r}^{*}}^{(l)}$ over RAT $l$ with a $k$-sparse $\bm{y}^{(l)}$. For each RAT $l$, let us begin with a generally nonsparse optimal solution $\bm{y}^{(l)}$, whose support is $\mathcal{S}^{(l)}$ ($y^{\mathcal{A},l} = 0$ if $\mathcal{A} \notin \mathcal{S}^{(l)}$). Let us freeze the optimal $z_{i \rightarrow j}^{\mathcal{A},l}$ variables and define a $k$-vector $\bm{q}^{\mathcal{A},l}$ for every $\mathcal{A} \in \mathcal{S}^{(l)}$ with its elements determined by $q_{j}^{\mathcal{A},l}=\sum_{i \in \mathcal{N}}s_{i \rightarrow j}^{\mathcal{A},l}z_{i \rightarrow j}^{\mathcal{A},l}$. According to (P4a), a convex combination of the vectors $(\bm{q}^{\mathcal{A},l})_{\mathcal{A} \in \mathcal{S}^{(l)}}$ with $(y^{\mathcal{A},l})_{\mathcal{A} \in \mathcal{S}_{(l)}}$ as coefficients form the optimal rate tuple $\bm{r}^{(l)}$ over RAT $l$. By Carath$\text{\`{e}}$odory's Theorem \cite{Egg:Convexity}, $\bm{r}^{(l)}$ can be represented as a convex combination
	of at most $k + 1$ of those vectors. Moreover, $\bm{r}^{(l)}$ must be on the boundary, not in the interior of the convex hull of $(\bm{q}^{\mathcal{A},l})$, because otherwise the rate tuple can be increased in all dimensions, contradicting the optimality assumption. Thus, for each RAT $l$, we can identify a vector ${\bm{y}^*}^{(l)}$ whose support is a subset of $\mathcal{S}^{(l)}$ with $k$ or fewer elements, such that the optimal $\bm{r}^{(l)}$ is a convex combination of $(\bm{q}^{\mathcal{A},l})$ with ${y^*}^{\mathcal{A},l}$ as coefficients. \hfill $\blacksquare$
	%%%Appendix C
	\section{Proof of Theorem 2}
	
		\begin{figure}[t]
			\centering
			\includegraphics[width=5in]{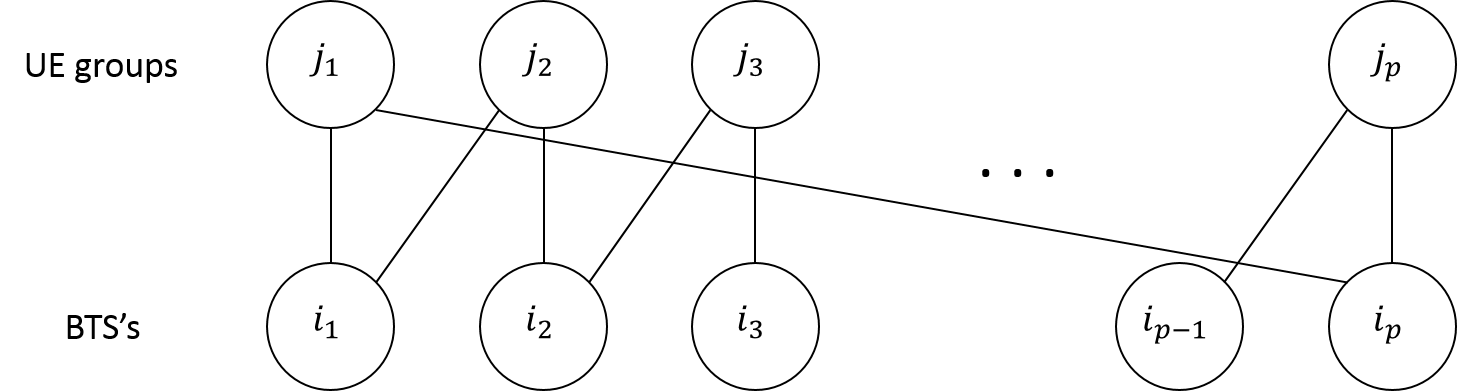}
			\caption{A loop in the BGR for the UE groups served by multiple APs over the same RAT.}
			\label{BGR}
		\end{figure}
	
	The proof follows an analogous proof in \cite{Kuang:Optimal}. The KKT conditions for P1 for nonzero elements of $\bm{x}$ and $\bm{y}$ are :
	\begin{alignat*}{2}
	&\frac{\partial U}{\partial r^{(l)}_j}s_{i \rightarrow j}^{\mathcal{A},l}-\mu^{\mathcal{A},l}_i = 0, \tag{26a}\\
	&\sum_{i \in \mathcal{N}}\mu^{\mathcal{A},l}_i = \xi^{(l)}, \tag{26b}
	\end{alignat*}
	where $\mu^{\mathcal{A},l}_i$ and $\xi^{(l)}$ are the Lagrange multipliers for constraints (P1e) and (P1f), respectively. Assuming UE group $j$ is served by two APs $i_1$ and $i_2$ over RAT $l$, define $s_{i \rightarrow j}^{(l)} = \sum\limits_{\mathcal{A} \subset \mathcal{N}}s_{i \rightarrow j}^{\mathcal{A},l}$ and $\mu^{(l)}_i = \sum\limits_{\mathcal{A} \subset \mathcal{N}}\mu^{\mathcal{A},l}_i.$ According to (26), we have
	\begin{equation}
	\frac{s_{i_{1} \rightarrow j}^{(l)}}{s_{i_{2} \rightarrow j}^{(l)}}=\frac{\mu^{(l)}_{i_1}}{\mu^{(l)}_{i_2}}. \tag{27}\\
	\end{equation}
	Following the argument in \cite{Kuang:Optimal}, a bipartite graph representation is used as
	in \cite{Gajic:Competition}. For each RAT, denote the UE groups served by multiple APs and the corresponding APs as nodes. An edge between a UE group and an AP represents an association. It remains to show that the graph contains no loop. Suppose there is a loop in the bipartite graph as shown in Fig. 10. Then there exists a sequence of nodes of users $j_1,\ldots,j_p$ and a sequence of APs $i_1,\ldots,i_p$, where user $j_q$ is connected with AP $i_q$ for $q = 1,\ldots,p$, user $j_{q+1}$ is connected with AP $i_q$ for $q = 1,\ldots,p-1$, and user 1 is connected with AP $p$. The nodes are distinct otherwise we can find a smaller loop with this property. According to (27), the loop implies:
	\begin{displaymath}
	\frac{s_{i_{p} \rightarrow j_1}^{(l)}}{s_{i_{1} \rightarrow j_1}^{(l)}} \frac{s_{i_{1} \rightarrow j_2}^{(l)}}{s_{i_{2} \rightarrow j_2}^{(l)}} \ldots \frac{s_{i_{p-1} \rightarrow j_p}^{(l)}}{s_{i_{p} \rightarrow j_p}^{(l)}} = \frac{\mu^{(l)}_{i_p}}{\mu^{(l)}_{i_1}} \frac{\mu^{(l)}_{i_1}}{\mu^{(l)}_{i_2}} \ldots \frac{\mu^{(l)}_{i_{p-1}}}{\mu^{(l)}_{i_p}} = 1.
	\end{displaymath}
	Since $s_{i \rightarrow j}^{(l)}$ is the sum of spectral efficiencies over all patterns over RAT $l$, it is a random variable based on random topology. Therefore, $\frac{s_{i_{p} \rightarrow j_1}^{(l)}}{s_{i_{1} \rightarrow j_1}^{(l)}} \frac{s_{i_{1} \rightarrow j_2}^{(l)}}{s_{i_{2} \rightarrow j_2}^{(l)}} \ldots \frac{s_{i_{p-1} \rightarrow j_p}^{(l)}}{s_{i_{p} \rightarrow j_p}^{(l)}} = 1$ is a zero probability event, which shows that w.p.1 there is no loop in the graph. Since there are $n$ APs, the largest possible BGR without a loop has $n-1$ user nodes, which proves Theorem 2. \hfill $\blacksquare$

	%%%Appendix D
	\section{Proof of Proposition 2}
	Taking the second derivatives of $\lambda^{(1)}_j\hat{t}^{(1)}_j$ with respect to $\lambda^{(1)}_j$ and $r^{(1)}_j$, we have:
	\begin{alignat*}{4}
	&\frac{\partial^2\lambda^{(1)}_j\hat{t}^{(1)}_j}{\partial {\lambda_j^{(1)}}^2}=\frac{\eta r^{(1)}_j}{(r^{(1)}_j-\beta \lambda^{(1)}_j)^3},\tag{28a}\\
	&\frac{\partial^2\lambda^{(1)}_j\hat{t}^{(1)}_j}{\partial {r_j^{(1)}}^2}=\frac{\eta \lambda^{(1)}_j}{\beta}\left(\frac{1}{(r^{(1)}_j-\beta \lambda^{(1)}_j)^3}-\frac{1}{\left(r_j^{(1)}\right)^3}\right)+\frac{2\beta \lambda^{(1)}_j}{{\left(r_j^{(1)}\right)}^3}.\tag{28b}
	\end{alignat*}
	When $\frac{1}{\lambda^{(1)}_j}>\frac{\beta}{r^{(1)}_j}$, in other words, when the queue is stable, the derivatives are positive, which means $\lambda^{(1)}_j\hat{t}^{(1)}_j$ is bi-convex in $\lambda^{(1)}_j$ and $r^{(1)}_j$. In addition, since $\lambda^{(2)}_j\hat{t}^{(2)}_j$ is similar to $\lambda^{(1)}_j\hat{t}^{(1)}_j$ except for an additional term $\frac{\nu_j {(\lambda_j^{(2)})}^2 r^{(2)}_j}{2(r^{(2)}_j-\beta \lambda^{(2)}_j)^+}$, which is also bi-convex in $\lambda^{(2)}_j$ and $r^{(2)}_j$, $\lambda^{(2)}_j\hat{t}^{(2)}_j$ is also bi-convex in $\lambda^{(2)}_j$ and $r^{(2)}_j$. Since $\hat{U}$ is a linear combination of bi-convex functions, we conclude that (17) is bi-convex in $\bm{\lambda}$ and $\bm{r}$. \hfill $\blacksquare$
	
\end{appendices}



%\bibliographystyle{IEEEtran}

\end{document}